\documentclass[
aps,
prd,
reprint,
longbibliography,
floatfix,
superscriptaddress
]{revtex4-2}

\usepackage{amsmath}
\usepackage{amsfonts}
\usepackage{color}
\usepackage{graphicx}
\usepackage{hyperref}
\usepackage{capt-of}

\makeatletter
\AtBeginDocument{%
  \immediate\write\@auxout{\string\citation{apsrev42LongBibliography}}%
}
\makeatother

\begin{document}

\title{Medium effect on spin alignment of strange and charm vector mesons}

\author{Ruixiang Chen}
\email{chenruixiang22@mails.ucas.ac.cn}
\affiliation{School of Nuclear Science and Technology, University of Chinese Academy of Sciences, Beijing 101408, China}

\author{Hiwa A. Ahmed}
\email{hiwa.ahmed@chu.edu.iq}
\affiliation{Charmo Center for Research, Training, and Consultancy, Charmo University, 46023, Chamchamal, Sulaymaniyah, Iraq}

\author{Yidian Chen}
\email{chenyidian@hznu.edu.cn}
\affiliation{School of Physics, Hangzhou Normal University, Hangzhou, 311121, China}

\author{Mei Huang}
\email{huangmei@ucas.ac.cn}
\affiliation{School of Nuclear Science and Technology, University of Chinese Academy of Sciences, Beijing 101408, China}

\begin{abstract}
Understanding the spin alignment of vector mesons in relativistic heavy-ion collisions requires a nonperturbative description of their spin-dependent in-medium properties. We investigate this problem within a unified four-flavor soft-wall holographic framework that combines an anisotropic Einstein--Maxwell--dilaton background at finite temperature, baryon chemical potential, and angular velocity. Spin alignment is determined from the medium-induced splitting of the spin-resolved vector-current spectral functions through an instantaneous freeze-out prescription. We systematically study the strange and charm vector mesons $K^{*}$, $\phi$, $D^{*}$, $D_s^{*}$, and $J/\psi$ and the dependence of their spin alignment on transverse momentum, rapidity, temperature, baryon chemical potential, and angular velocity. We find that the heavy charm vector mesons $D^{*}$, $D_s^{*}$, and $J/\psi$ mesons exhibit $\rho_{00}>1/3$ at low transverse momentum, whereas the light strange vector mesons $K^{*}$ and $\phi$ exhibit the opposite low-momentum behavior and angular distributions with $\rho_{00}<1/3$. We trace this flavor-dependent separation to the different locations of the vacuum mass shell relative to the thermally shifted longitudinal and transverse spectral peaks. The results qualitatively reproduce several trends observed at low and intermediate transverse momentum. Spin alignment is insensitive to baryon chemical potential and only weakly affected by angular velocity. These results establish an equilibrium holographic baseline for vector-meson spin alignment across flavor sectors and help delineate the regimes in which additional mechanisms, such as nonequilibrium evolution, fluctuations, and hard production, become important.
\end{abstract}

\maketitle

\section{Introduction}

Relativistic heavy-ion collisions provide a unique laboratory for studying strongly interacting matter under extreme conditions. In noncentral collisions, the system carries a large orbital angular momentum, part of which is transferred to the quark-gluon plasma (QGP) produced in the collision. Through strong-interaction spin-orbit coupling, this angular momentum polarizes the constituent quarks and antiquarks, making the QGP the most vortical fluid observed so far. This picture of global polarization~\cite{Liang:2004ph,Becattini:2020ngo} is supported experimentally by measurements of $\Lambda$-hyperon polarization~\cite{STAR:2017slg}. For vector mesons, the hadronization of polarized quarks and antiquarks populates the three spin states unequally, leading to spin alignment characterized by the spin-density-matrix element $\rho_{00}$, which deviates from the unpolarized value of $1/3$~\cite{Liang:2004xn}. Because $\rho_{00}$ can be extracted directly from the angular distribution of the meson decay products, it provides a sensitive probe of spin dynamics in the strongly interacting medium.

Experimentally, evidence for nonzero spin alignment of $K^{*0}$ and $\phi$ mesons was first reported by ALICE in Pb--Pb collisions~\cite{ALICE:2019aid}. Later, STAR established significant global spin alignment of the $\phi$ meson, with deviations of $\rho_{00}$ from $1/3$ far exceeding the theoretical estimates available at the time, while the $K^{*0}$ result remained consistent with $1/3$~\cite{STAR:2022fan}. More recently, measurements have been extended to $J/\psi$ polarization~\cite{ALICE:2022dyy} and $D^{*+}$ spin alignment~\cite{ALICE:2025cdf}. These measurements demonstrate the relevance of spin alignment across a broad range of meson species and collision energies.

Understanding the origin of these observations remains an active theoretical challenge. Within the hydrodynamic-statistical framework, spin polarization arises from thermal vorticity and thermal shear in the medium~\cite{Becattini:2007sr,Becattini:2013fla,Becattini:2021iol}. Although this framework successfully describes the global polarization of hyperons, it predicts deviations of vector-meson $\rho_{00}$ from $1/3$ that are much smaller than the measured values, and the coalescence of vorticity-polarized quarks leads to similarly small deviations~\cite{Liang:2004xn,Yang:2017sdk}. A key observation is that, unlike hyperon polarization, which reflects the spacetime-averaged quark polarization, vector-meson spin alignment is also sensitive to local correlations between the spin polarizations of the quark and antiquark~\cite{Sheng:2022wsy,Xin-Li:2023gwh}. This feature leaves room for fluctuation-driven mechanisms. For example, fluctuations of the strong-force field, described in terms of an effective vector field, can generate sizable spin alignment through correlated quark-antiquark polarization~\cite{Sheng:2019kmk,Sheng:2022wsy,Sheng:2022ffb,Sheng:2023urn}, while turbulent color fields in the early stage of the collision may produce a comparable effect~\cite{Muller:2021hpe,Kumar:2023ghs}. Other proposed mechanisms include helicity polarization of vector mesons~\cite{Gao:2021rom}, magnetic-field-induced mass splitting~\cite{Sheng:2022ssp}, kinetic and linear-response descriptions of spin dynamics~\cite{Li:2022vmb,Dong:2023cng,Fang:2023bbw}, and direct coupling between the meson spin and medium rotation~\cite{WeiMingHua:2020eee,Wei:2023pdf,Sun:2024anu}. Despite their diversity, these mechanisms attribute spin alignment to the spin-dependent in-medium properties of the mesons or their constituents, which are governed by nonperturbative strong-coupling dynamics and therefore call for genuinely nonperturbative tools.

The gauge/gravity duality~\cite{Maldacena:1997re,Gubser:1998bc,Witten:1998qj,Aharony:1999ti} provides one of the most powerful nonperturbative frameworks for studying strongly coupled gauge theories. In the large-$N_c$ limit, certain strongly coupled gauge theories are dual to classical gravity in a higher-dimensional, asymptotically anti-de Sitter spacetime. Within this correspondence, thermal states of the boundary theory are mapped onto black-hole geometries in the bulk~\cite{Witten:1998zw}, while real-time correlation functions of boundary operators can be extracted from the classical dynamics of the corresponding bulk fields~\cite{Son:2002sd}.

Applied to QCD, holography has developed into a versatile framework for investigating hadron physics, hot and dense QCD matter, and real-time transport phenomena. In hadron physics, both top-down constructions~\cite{Sakai:2004cn} and phenomenological bottom-up models~\cite{Erlich:2005qh,Karch:2006pv,Gherghetta:2009ac,Fang:2016nfj} successfully reproduce Regge-like meson spectra and describe chiral symmetry breaking. These models have also been extended to the strange and charm sectors~\cite{Abidin:2009aj,Chen:2021wzj}. At finite temperature and density, holographic models reproduce the equation of state obtained from lattice QCD and provide realistic descriptions of the QCD phase structure~\cite{Gursoy:2007er,Gubser:2008yx,DeWolfe:2010he,Dudal:2017max,Chen:2024mmd}, while also yielding important insights into transport properties such as shear viscosity and the jet-quenching parameter~\cite{Policastro:2001yc,Kovtun:2004de,Liu:2006ug}. The same framework naturally provides access to real-time observables, including in-medium spectral functions and the thermal dissociation of mesons~\cite{Cao:2021tcr}, and has recently been extended to rotating strongly interacting matter~\cite{Braguta:2021jgn,Chen:2022mhf,Chen:2024jet,Sheng:2026iiw}.

These developments make holography a natural framework for investigating vector-meson spin alignment. A holographic formalism has been proposed in which spin alignment is determined jointly by vector-current spectral functions projected onto definite spin states and an instantaneous freeze-out prescription~\cite{Sheng:2024kgg}. Within this framework, deviations of $\rho_{00}$ from $1/3$ originate from the medium-induced splitting of the spectral functions associated with different polarization states. The formalism has subsequently been applied to the $J/\psi$ meson in magnetized matter~\cite{Zhao:2024ipr}, and the effects of gluon polarization have also been investigated holographically~\cite{Ahmed:2025bwi}.

Building on these developments, we extend the holographic description of spin alignment to vector mesons across different flavor sectors. Using a unified bottom-up framework incorporating finite temperature, baryon chemical potential, and rotation, we investigate the spin alignment of the $K^{*}$, $\phi$, $D^{*}$, $D_s^{*}$, and $J/\psi$ mesons. We also analyze its dependence on transverse momentum, rapidity, temperature, chemical potential, and angular velocity, and systematically compare the results with available experimental measurements.

The remainder of this paper is organized as follows. Sec.~\ref{sec:Holographic Model} introduces the holographic model, including the anisotropic rotating background and the vector-meson matter sector. The retarded correlation functions, quasinormal-mode analysis, and resulting meson spectrum and spectral functions are presented in Sec.~\ref{sec:Spectral Function and Mass spectrum}. Sec.~\ref{sec:Spin Alignment} presents the spin-alignment formalism and numerical results, organized around their flavor dependence, comparison with experimental data, and physical interpretation. Detailed results for the individual meson species are collected in Appendix~\ref{app:parameter dependence}. Sec.~\ref{sec:conclusion} summarizes our conclusions.

\section{Holographic Model}
\label{sec:Holographic Model}

The holographic approach is based on the following AdS/CFT dictionary:
\begin{equation}
    \left\langle e^{i\,\,\int{d^4xX_0\left( x \right) O_G\left( x \right)}} \right\rangle _{4D}=e^{iS_{5D}\left[ X\left( x,z \right) \right]},
\label{eq:GKB}
\end{equation}
where $S_{5D}\left[ X\left( x,z \right) \right]$ is the on-shell five-dimensional action of the bulk field $X(x,z)$ dual to $O_G\left( x \right)$, and $X_0(x)=X(x,0)$ is the boundary source. Correlation functions in QCD can therefore be obtained by differentiating the right-hand side of Eq.~\eqref{eq:GKB} with respect to the source. The five-dimensional action consists of two parts:
\begin{equation}
    S_{5D}=S_B+S_M,
\end{equation}
where $S_B$ is the background action describing a hot and dense medium, while $S_M$ is the matter-field action describing vector mesons. We adopt the probe approximation and neglect the backreaction of the matter fields on the background.

We employ the Einstein--Maxwell--dilaton (EMD) system with an anisotropic metric~\cite{Chen:2024jet} as the background action and a four-flavor extension of the soft-wall model~\cite{Chen:2021wzj} as the matter-field action.

\subsection{The Einstein--Maxwell--dilaton model under rotation}

The EMD action in the string frame is
\begin{equation}
\begin{aligned}
    S_B ={}& \frac{1}{16\pi G_5} \int d^5x \sqrt{-g_s} e^{-2\Phi} \\
    &\quad{}\times\Bigg[ R_s + 4\partial_M \Phi \partial^M \Phi - V_s (\Phi) \\
    &\qquad{}- \frac{h(\Phi)}{4} e^{\frac{4\Phi}{3}} F_{MN} F^{MN} \Bigg],
\end{aligned}
\end{equation}
where $G_5$ is the five-dimensional Newton constant and the subscript $s$ denotes the string frame. The scalar field $\Phi$ is the dilaton associated with the gluonic dynamics, and $V_s(\Phi)$ is its potential. The tensor $F_{MN}$ is the field strength of the $U(1)$ gauge field $A_M$ associated with the baryon-number current.

The thermodynamics of QCD matter can be reasonably well described by an EMD system whose metric admits a black-hole solution. This analysis is usually carried out in the Einstein frame. The two frames are related to each other through the following conformal transformation:
\begin{equation}
\begin{aligned}
    g^s_{MN}&=e^{\frac{4}{3}\Phi}g^e_{MN},\qquad
    \phi=\sqrt{\frac{8}{3}}\Phi,\\
    V_e(\phi)&=e^{\frac{4}{3}\Phi}V_s(\Phi).
\end{aligned}
\end{equation}
The action in the Einstein frame is then given by
\begin{equation}
\begin{aligned}
    S_B ={}& \frac{1}{16\pi G_5} \int d^5x \sqrt{-g_e}
    \Bigg[ R_e - \frac{1}{2} \partial_M \phi \partial^M \phi - V_e (\phi) \\
    &\qquad{}- \frac{h(\phi)}{4} F_{MN} F^{MN}\Bigg].
\end{aligned}
\end{equation}

This action describes QCD matter at finite temperature and density. Because the large angular momentum generated in heavy-ion collisions can affect spin alignment through mechanisms such as spin-orbit coupling, we extend the background to finite angular velocity. The construction developed in Ref.~\cite{Chen:2024jet} captures rotational effects on the deconfinement and chiral phase transitions in a manner consistent with recent lattice QCD results~\cite{Braguta:2021jgn}. Following this approach, we consider the anisotropic background

\begin{equation}
\label{eq:metric}
\begin{aligned}
    ds^2 ={}& \frac{e^{2A_e(z)}}{z^2} \Bigg[ -f(z) dt^2 + \frac{dz^2}{f(z)}
    + e^{B(z)} dr^2 \\
    &\qquad{}+ r^2 e^{B(z)} d\theta^2 + e^{-2B(z)} dx_3^2 \Bigg],
\end{aligned}
\end{equation}
where $r$ and $\theta$ are the radial and angular coordinates, respectively, and $x_3$ runs along the symmetry axis. The blackening factor $f(z)$ satisfies $f(z_h)=0$ at the black-hole horizon $z_h$. Asymptotically AdS behavior near the boundary $z=0$ further requires $f(0)=1$, $A_e(0)=0$, and $B(0)=0$.

Moreover, because the temporal and azimuthal components of the baryon-number current $\bar{q}\gamma^{\mu}q$ are expected to be nonzero in a rotating system, we choose the gauge-field configuration $A_{M}=(A_{t},0,A_{\theta},0,0)$ in the $A_z=0$ gauge. The boundary value of $A_t$ is the chemical potential and serves as the source for the operator $\bar{q}\gamma^{0}q$, so that $A_t(0)=\mu$. As in Ref.~\cite{Chen:2022mhf}, $A_\theta$ can be expanded as $A_\theta \sim \Omega r^2 + \rho(r,z)$. Within the near-center approximation, we therefore set $A_\theta=\Omega r^2$.

This setup yields the following equations of motion:

\begin{subequations}

\begin{multline}
    \frac{z^2 h(\phi) e^{-2A_e} (A_t^{\prime 2} + 4e^{-2B}\Omega^2)}{4f}
    + \frac{3f^\prime}{2f} \left( A_e^\prime - \frac{1}{z} \right) \\
    + \frac{e^{2A_e} V_e(\phi)}{2z^2 f} - \frac{12A_e^\prime}{z}
    + 6A_e^{\prime 2} \\
    - \frac{3B^{\prime 2}}{4} - \frac{2\phi^{\prime 2}}{3}
    + \frac{6}{z^2} = 0,
\end{multline}

\begin{equation}
\begin{aligned}
    &-z^2 e^{-2A_e} h(\phi)
    \left( A_t'^2 + \frac{8}{3} e^{-2B}\Omega^2 \right) \\
    &\qquad{}+ 3f' \left( A_e' - \frac{1}{z} \right) + f'' = 0,
\end{aligned}
\end{equation}

\begin{equation}
    \frac{B'^2}{2} + \frac{4\Phi'^2}{9} + \frac{2A_e'}{z} - A_e'^2 + A_e'' = 0,
\end{equation}

\begin{equation}
     \frac{4\Omega^2 z^2 h(\phi) e^{-2(A_e+B)}}{3f} + B' \left( 3A_e' + \frac{f'}{f} - \frac{3}{z} \right) + B'' = 0,
\end{equation}

\begin{equation}
    A_t' \left( A_e' + \frac{h'}{h} - \frac{1}{z} \right) + A_t'' = 0,
\end{equation}

\begin{equation}
\begin{aligned}
    &z^2 e^{-2A_e} \partial_{\phi} h(\phi)
    (A_t'^2 - 4e^{-2B}\Omega^2)
    - \frac{e^{2A_e} \partial_{\phi} V_e(\phi)}{z^2 f} \\
    &\qquad{}+ \phi' \left( 3A_e' + \frac{f'}{f} - \frac{3}{z} \right)
    + \phi'' = 0.
\end{aligned}
\end{equation}

\end{subequations}

\begin{figure*}[!t]
    \centering
    \includegraphics[width=0.95\textwidth]{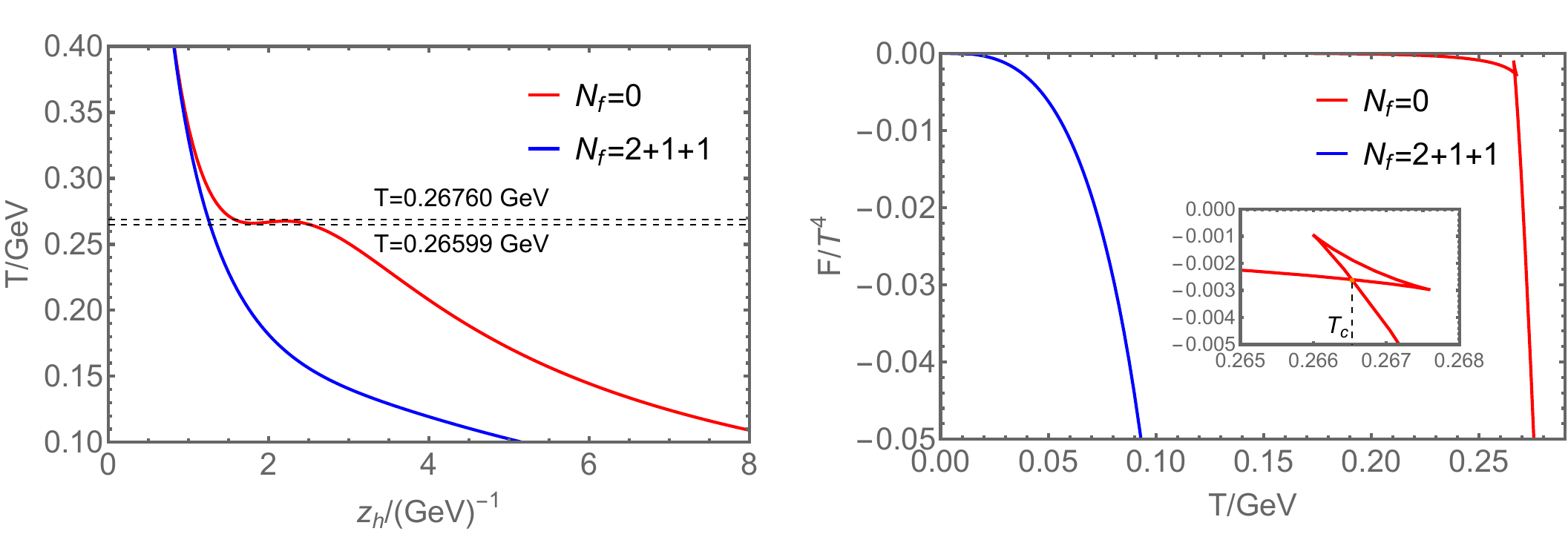}
    \caption{Left panel: the relationship between the temperature and the horizon position $z_h$. Right panel: the free energy density as a function of temperature. The red line corresponds to the pure gluon system, with parameters $a=0$, $b=0.072$, $c=0$, $d=-0.584$, $k=0$, $G_5=1.326$~\cite{Chen:2024mmd}. The blue line corresponds to the $N_f=2+1+1$ system, with parameter values given in the text.}
    \label{fig:T-zh&F-T}
\end{figure*}

To solve this set of equations, we adopt the potential-reconstruction method, taking $A_e$ and $h(\phi)$ as inputs and reconstructing $V(\phi)$ from the resulting background solutions. We choose the following warp factor:
\begin{equation}
\label{eq:input Ae}
    A_e(z,\Omega)=d \, \text{log}[(a+w\Omega^2)z^2 + 1] + d \, \text{log}[(b+w^2 \Omega^4)z^4 + 1],
\end{equation}
with four free parameters $a$, $b$, $d$, and $w$. This form admits black-hole solutions at arbitrary temperatures and, depending on the parameter choice, realizes either a first-order deconfinement transition for a pure-gluon system or a crossover for $N_f \geq 2$~\cite{Dudal:2017max,Chen:2024mmd}. We include terms involving $w\Omega^2$ to account for rotational effects. Because we use $A_e(z)$ instead of $\phi(z)$ as the input, this parametrization differs from that of Ref.~\cite{Chen:2024jet}, where the rotational correction is introduced through $\phi(z)$. Owing to the coupling between the dilaton and warp factor in the EoMs, the parametrization in Eq.~\eqref{eq:input Ae}, once fitted to lattice results, can be regarded as an equivalent effective description of the rotational modification of the gluon dynamics. For the coupling between the dilaton and gauge field, we set
\begin{equation}
    h(\phi(z)) = e^{cz^2 - A_e(z)+k},
\end{equation}
where $c$ and $k$ are undetermined parameters.

These ingredients determine the background fields and their thermodynamic properties. The model parameters are fixed by fitting the resulting QCD equation of state to lattice-QCD data.

\subsubsection{Thermodynamics of the 2+1+1-flavor system under rotation}

The thermodynamic properties of the strongly coupled matter are derived from the black-hole solution through the holographic dictionary. In the AdS/CFT correspondence, a black hole in the bulk corresponds to a thermal state in the boundary field theory, with the Hawking temperature and Bekenstein--Hawking entropy identified with the field-theory temperature and thermal entropy, respectively~\cite{Witten:1998zw,Chen:2024mmd}.

The temperature is determined by the surface gravity at the black-hole horizon $z=z_h$. For the metric ansatz in Eq.~\eqref{eq:metric}, the Hawking temperature is
\begin{equation}
    T = \frac{|\kappa|}{2\pi} = \frac{1}{4\pi}\left| \frac{df}{dz} \right|_{z=z_h}.
    \label{eq:temperature}
\end{equation}
The entropy density is obtained from the Bekenstein--Hawking area law,
\begin{equation}
    s = \frac{A_{\text{area}}}{4G_5} \Big|_{z=z_h}
      = \frac{e^{3A_e(z_h)}}{4G_5\, z_h^{3}}.
    \label{eq:entropy}
\end{equation}
The free-energy density $F$ follows from the thermodynamic relation in the grand canonical ensemble,
\begin{equation}
    dF = -s\,dT - \rho\,d\mu - j\,d\Omega,
    \label{eq:df}
\end{equation}
where $\rho$ is the baryon number density and $j$ is the angular momentum density. Integrating along a path of fixed chemical potential $\mu$ and fixed angular velocity
$\Omega$, one obtains
\begin{equation}
    F(T,\mu,\Omega) = \int_{z_{h}(T)}^{\infty} s(T',\mu,\Omega)\,\frac{dT'}{dz_h}dz_h,
    \label{eq:free_energy}
\end{equation}
with the boundary condition $F\to 0$ as $T\to 0$ (or $z_h\to\infty$), where the black hole vanishes and the system reduces to the vacuum. 

Several other thermodynamic quantities can be derived from the same black-hole solution, including the energy density $\epsilon=F+Ts+\mu \rho+\Omega j$, the pressure $P=-F$, and the squared speed of sound $C_s^2 = \left. \frac{dP}{d\epsilon} \right|_{\mu,\,\Omega}$. Reference~\cite{Chen:2024mmd} used the same background ansatz without rotation and employed a deep-learning method to fit lattice-QCD results for the equation of state at zero chemical potential. Parameters were obtained for the pure-gluon system and for different numbers of flavors. For $N_f=2+1+1$, they are $a = 0.196,\, b = 0.014,\, c = -0.362,\, d = -0.171,\, k = -0.735$, and $G_5 = 0.391$.

Using these parameters, we plot the temperature as a function of the horizon position, $T(z_h)$, and the free energy as a function of temperature, $F(T)$, in Fig.~\ref{fig:T-zh&F-T}. We also show the pure-gluon case, which will be used below to fit the angular-velocity dependence of the deconfinement transition temperature. For the pure-gluon system, $T(z_h)$ is nonmonotonic: a monotonically increasing branch is flanked by two monotonically decreasing branches. Correspondingly, $F(T)$ develops a swallowtail structure. In the multivalued region of $F(T)$, the black-hole branch with the lowest free energy is thermodynamically favored. The temperature at which the two stable branches intersect therefore defines a first-order transition. For $N_f=2+1+1$, both $T(z_h)$ and $F(T)$ are monotonic, indicating a crossover.

\begin{figure}[!t]
    \centering
    \includegraphics[width=0.9\linewidth]{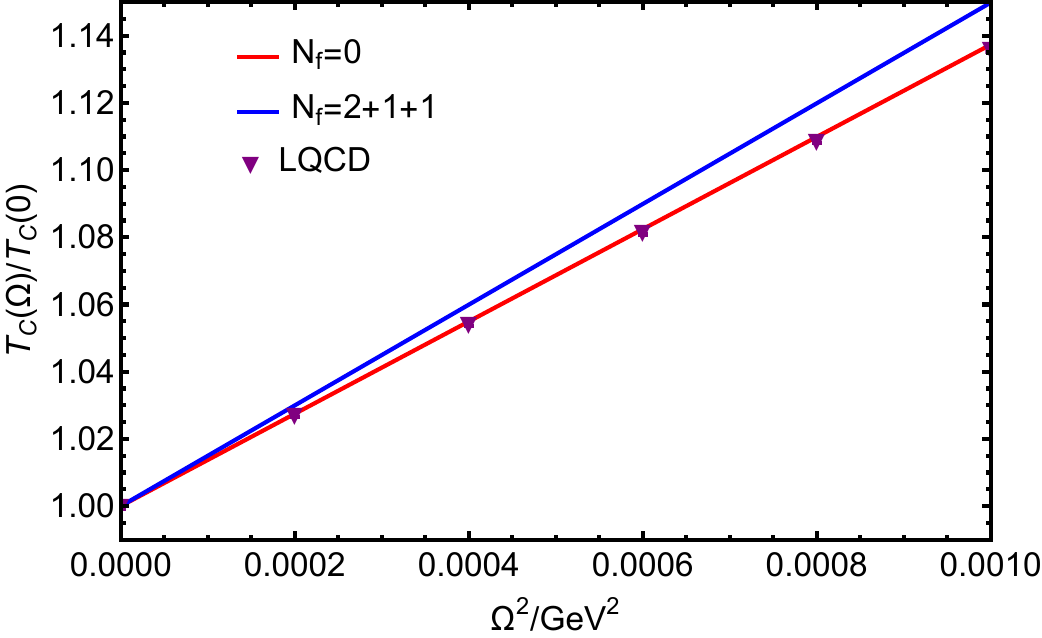}
    \caption{The ratio of the deconfinement transition temperature \(T_c(\Omega)\) at angular velocity \(\Omega\) to that without rotation \(T_c(0)\). The red line corresponds to the pure gluon case, obtained by fitting the lattice results~\cite{Braguta:2021jgn} shown in the figure. The blue line corresponds to the \(N_f=2+1+1\) case, using the fitted parameter \(w=92.08\).}
    \label{fig:Tc-omega}
\end{figure}

With these parameters fixed, we determine the rotation-dependent parameter $w$ by fitting the deconfinement transition temperature $T_c(\Omega)$ to the lattice results~\cite{Braguta:2021jgn}. Because these results were obtained for a pure-gluon system, we use the corresponding model parameters and determine the transition temperature at each angular velocity following the procedure described above. The results are shown in Fig.~\ref{fig:Tc-omega}. The lattice data are described by $T_c(\Omega)/T_c(0)=1+C\Omega^2$. For the pure-gluon system, the fit yields $w=92.08$ and $C=137.34\,\mathrm{GeV}^{-2}$. We then use the fitted value of $w$ to calculate $T_c(\Omega)$ for $N_f=2+1+1$. In this case, the inflection point of the squared speed of sound $C_s^2(T)$ is used to define the crossover temperature; further details are given in Ref.~\cite{Chen:2024mmd}. The resulting values are $T_c(0)=129.42\,\mathrm{MeV}$ and $C=149.84\,\mathrm{GeV}^{-2}$.

Ideally, the angular-velocity dependence of $T_c$ for $N_f=2+1+1$ should be fitted directly to lattice data, because dynamical quarks may modify this dependence. Reliable lattice results for this system are not yet available. In our model, quark dynamics is described primarily by the matter action $S_M$, whereas the background action $S_B$ mainly encodes gluon dynamics under the probe approximation. Nevertheless, choosing flavor-dependent background parameters that reproduce the corresponding equations of state incorporates quark effects indirectly. We expect this treatment to capture the leading change in the rotational response induced by dynamical quarks.

\subsection{Vector-meson multiplets in the holographic model}
\label{sec:matter part}

In this section, we introduce a holographic model for vector-meson multiplets including heavy flavors. This model constitutes the matter action $S_M$ and is inspired by the two-flavor hard-wall~\cite{Erlich:2005qh} and soft-wall~\cite{Karch:2006pv} models, as well as by subsequent extensions to $N_f>2$~\cite{Abidin:2009aj,Chen:2021wzj}.

According to the AdS/CFT correspondence, each operator in the 4D field theory corresponds to a field in AdS$_5$ space, and the five-dimensional mass of the field can be related to the conformal dimension of the corresponding $q$-form operator through the following formula:

\begin{equation}
    M_{5}^{2}=\left( \varDelta -q \right) \left( \varDelta +q-4 \right).
\end{equation}

\noindent Therefore, the five-dimensional fields dual to the operators relevant to chiral dynamics are listed in Tab.~\ref{tab:O/F}. Here, $t^a$ ($a=1,2,\ldots,N_f^2-1$) are the generators of $\mathrm{SU}(N_f)$, normalized according to $\mathrm{Tr}(t^a t^b)=\frac{1}{2}\delta^{ab}$.

\begin{table}
\begin{ruledtabular}
\begin{tabular}{ccccc}
4D Operator & 5D Field & $q$ & $\Delta$ & $m_5^2$\\ \hline
$\bar{q}_L \gamma^{\mu} t^a q_L$ & $A^{\mu}_L$ & 1 & 3 & 0 \\
$\bar{q}_R \gamma^{\mu} t^a q_R$ & $A^{\mu}_R$ & 1 & 3 & 0 \\
$\bar{q}_L q_R$ & X & 0 & 3 & -3
\end{tabular}
\end{ruledtabular}
\caption{\label{tab:O/F}Five-dimensional fields corresponding to four-dimensional $q$-form operators, where $\Delta$ is the conformal dimension of the operator and $m_5$ is the five-dimensional mass of the field.}
\end{table}

By promoting the chiral symmetry $SU(N_f)_L \times SU(N_f)_R$ to a gauge symmetry and introducing the dilaton field $\Phi_m$ to break conformal invariance, the following five-dimensional action can be constructed:

\begin{equation}
\label{eq:action}
\begin{aligned}
    S_M={}&-\int{d^5x}\sqrt{g}e^{-\Phi_{m}}\mathrm{Tr}\Bigg[ \\
    &\qquad{}\left| DX \right|^2+m_{5}^{2}(z)\left|X\right|^2
    -\kappa \left|X\right|^4 \\
    &\qquad{}+ \frac{1}{4g_{5}^{2}}\left( F_{L}^{2}+F_{R}^{2} \right) \Bigg],
\end{aligned}
\end{equation}

\noindent where $g_5$ and $\kappa$ are undetermined parameters. The field strength is $F_{MN}^{L/R}=\partial _MA_{N}^{L/R}-\partial _NA_{M}^{L/R}-i\left[ A_{M}^{L/R},A_{N}^{L/R} \right]$, and the covariant derivative acting on the scalar field $X$ is given by $D_MX=\partial _MX-iA_{M}^{L}X+iXA_{M}^{R}$. Moreover, we add a $z$-dependent correction to the original five-dimensional mass of the scalar field, so that $m_{5}^{2}(z)=-3 + \delta m_{5}^{2}(z)$. The reason for this modification will be explained in Sec.~\ref{Sec:Vacuum Sector}.

Linear combinations of $A_L$ and $A_R$ define the vector field $V^{\mu a} = \frac{1}{2}(A_L^{\mu a} + A_R^{\mu a})$ and the axial-vector field $A^{\mu a} = \frac{1}{2}(A_L^{\mu a} - A_R^{\mu a})$, which correspond to the vector and axial-vector operators $J_V^{\mu a}=\bar{q} \gamma^{\mu} t^a q$ and $J_A^{\mu a}=\bar{q} \gamma^{\mu} \gamma^5 t^a q$, respectively. Because we focus on vector mesons and neglect vector--axial-vector interactions, we omit the axial-vector field hereafter. The action then becomes
\begin{equation}
\label{eq:action*}
\begin{aligned}
    S={}&-\int{d^5x}\sqrt{g}e^{-\Phi_{m}} \mathrm{Tr}\Bigg[ \\
    &\qquad{}\left| DX \right|^2+m_{5}^{2}(z)\left|X\right|^2
    -\kappa \left|X\right|^4 \\
    &\qquad{}+ \frac{1}{2g_{5}^{2}} \tilde{V}^{MN}\tilde{V}_{MN} \Bigg],
\end{aligned}
\end{equation}
with
\begin{equation}
\begin{aligned}
    \tilde{V}_{MN}&=\partial _MV_{N}-\partial _NV_{M}-i[V_M,V_N],\\
    D_MX&=\partial _MX-i[V_{M},X].
\end{aligned}
\end{equation}
Since the present work includes charmed mesons, we set $N_f=4$. The correspondence between the components of the vector field and the meson states is summarized as follows:

\begin{center}
\(\displaystyle V = V^a t^a =\)\par\smallskip
\resizebox{0.98\columnwidth}{!}{$\displaystyle
\frac{1}{\sqrt{2}}
\begin{pmatrix}
\frac{\rho^0}{\sqrt{2}} + \frac{\omega'}{\sqrt{6}} + \frac{\psi}{\sqrt{12}} & \rho^+ & K^{*+} & \bar{D}^{*0} \\
\rho^- & -\frac{\rho^0}{\sqrt{2}} + \frac{\omega'}{\sqrt{6}} + \frac{\psi}{\sqrt{12}} & K^{*0} & D^{*-} \\
K^{*-} & \bar{K}^{*0} & -\sqrt{\frac{2}{3}}\omega' + \frac{\psi}{\sqrt{12}} & D_s^{*-} \\
D^{*0} & D^{*+} & D_s^{*+} & -\frac{3}{\sqrt{12}}\psi
\end{pmatrix}.
$}
\end{center}

For the coupling between the dilaton and matter fields, the original soft-wall model uses $e^{-\Phi_m(z)}$, with $\Phi_m(z)=\mu^2z^2$ and $\mu$ determined phenomenologically. This form is motivated by open-string couplings and yields a Regge-like meson spectrum. In many dynamical holographic models, $\Phi_m(z)$ is identified with the dilaton $\Phi(z)$ of the background EMD model. However, Ref.~\cite{Chen:2025ncu} showed that the model does not reproduce the observed meson spectrum when the warp factor $A_e$ in Eq.~\eqref{eq:input Ae} is used as the EMD input. Because the string scale lies far above the energy scale of interest, we instead adopt an effective coupling $g(\Phi)$ calibrated to experimental observables. Following Ref.~\cite{Chen:2025ncu}, we choose $g(\Phi(z))=e^{-\mu_g^2z^2}$, which is equivalent to setting $\Phi_m(z)=\mu_g^2z^2$.

\subsubsection{Vacuum sector}
\label{Sec:Vacuum Sector}

Because the scalar field $X$ is dual to the operator $\bar{q}_Lq_R$, we assign it a $z$-dependent vacuum expectation value (VEV),
\begin{equation}
    X=\frac{1}{2}\text{diag}[\chi_l(z),\chi_l(z),\chi_s(z),\chi_c(z)].
\end{equation}
Unlike the scalar field, the vector field $V$ acquires no VEV and is treated as a fluctuation about the vacuum solution $\big(X_0(z),V=0\big)$. We therefore expand the action in powers of $V$ as $S=S^{(0)}+S^{(2)}+S^{(3)}+\cdots$. The vacuum sector is described by the zeroth-order action

\begin{widetext}
\begin{equation}
\begin{aligned}
    S^{(0)}={}&\frac{1}{4}\int{d^5x}\Bigg\{
    \frac{e^{5A_s(z)-\Phi_m(z)}}{z^5}
    \Bigg[ -m_{5}^{2}(z)\left( 2\chi_l(z)^2+\chi_s(z)^2 + \chi_c(z)^2\right)
    +\frac{\kappa}{4}\left(2\chi_l(z)^4+\chi_s(z)^4+\chi_c(z)^4\right) \Bigg] \\
    &\qquad{}-\frac{e^{3A_s(z)-\Phi_m(z)}f(z)}{z^3}
    \Bigg[ 2\chi_{l}^{\prime}(z)\chi_{l}^{\prime}(z)
    +\chi_{s}^{\prime}(z)\chi_{s}^{\prime}(z)
    +\chi_{c}^{\prime}(z)\chi_{c}^{\prime}(z) \Bigg] \Bigg\}.
\end{aligned}
\end{equation}
\end{widetext}

Varying this action yields the equations of motion for $\chi_q(z)$, with $q=l,s,c$:

\begin{equation}
\label{eq:EoM of chi}
\begin{aligned}
    &\frac{z^3}{e^{5A_s-\Phi_m}}\partial _z\left[
    \frac{e^{3A_s-\Phi_m}f}{z^3}\partial _z\chi_q(z) \right] \\
    &\qquad{}-\frac{m_{5}^{2}(z)}{z^2}\chi_q(z)
    +\frac{\kappa}{2z^2}\chi_{q}^{3}(z)=0.
\end{aligned}
\end{equation}
Near the boundary $z\to0$, the field $\chi_q(z)$ has the asymptotic expansion
$\chi_q(z)=m_q\zeta z+\frac{\sigma_q}{\zeta}z^3+\cdots$, where $m_q$ is the quark mass, $\sigma_q$ is the chiral condensate according to the AdS/CFT dictionary, and $\zeta=\sqrt{N_c}/(2\pi)$ is a normalization constant~\cite{Cherman:2008eh}. We impose $m_q$ as a UV boundary condition, while regularity at the horizon $z_h$ fixes $\sigma_q$ and thereby determines $\chi_q(z)$.

\begin{figure}
    \centering
    \includegraphics[width=0.9\linewidth]{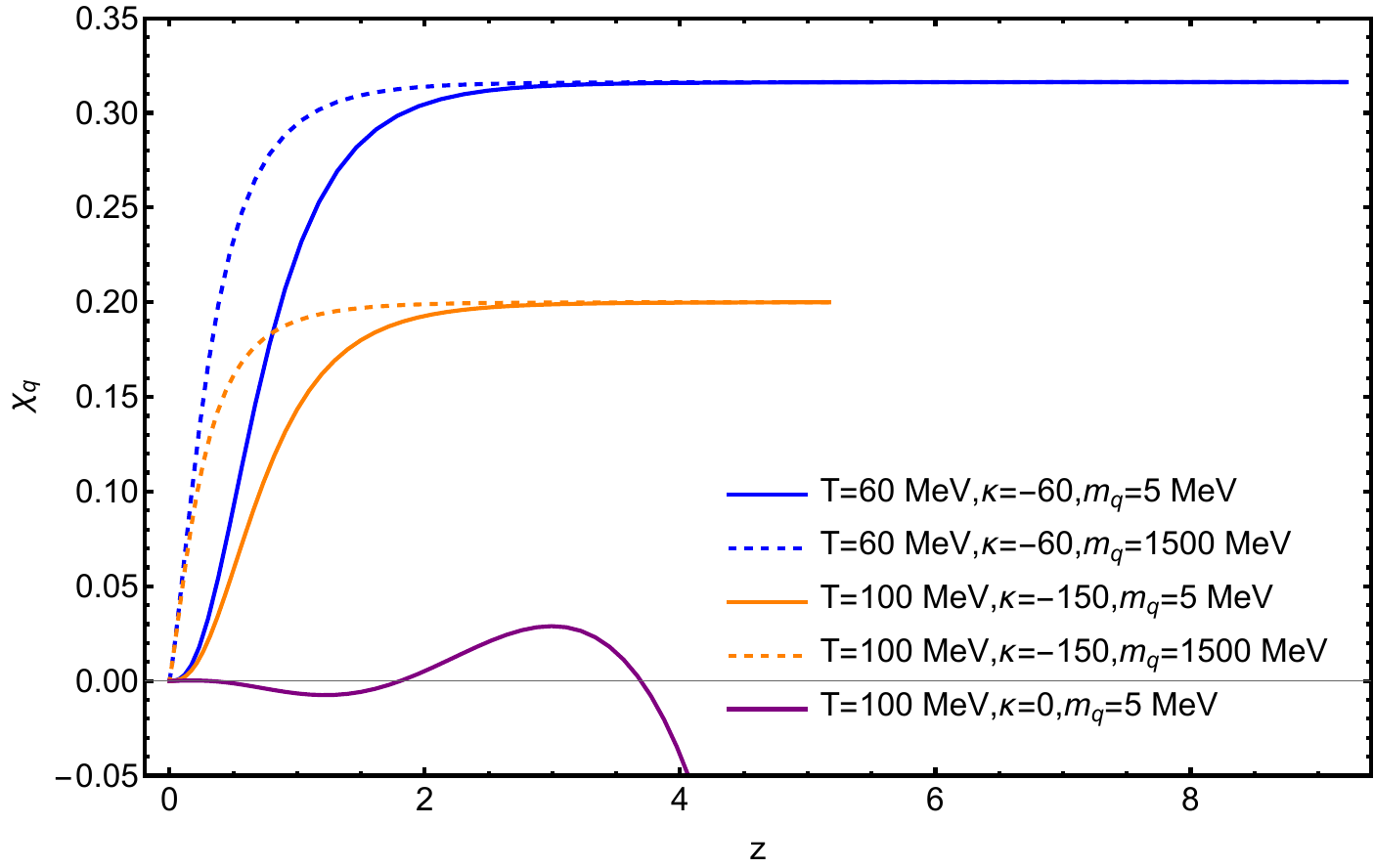}
    \caption{Scalar profiles $\chi_q$ obtained for different parameter values with $V(\chi_q)=m_5^2\chi_q^2-\kappa\chi_q^4$. Unless otherwise indicated, $\mu_g^2=0.205\,\mathrm{GeV}^2$ is used.}
    \label{fig:chi-z}
\end{figure}

We first solve the equation without the $z$-dependent correction to the five-dimensional mass, i.e., with $m_5^2=-3$. The result is shown in Fig.~\ref{fig:chi-z}. At large $z$, $\chi_q(z)$ approaches an approximately flavor-independent constant, while the effects of different quark masses appear only in the UV region ($z\sim0$). As shown by the expressions below, differences among the meson masses arise primarily from differences among the scalar profiles of their constituent quarks. The profiles obtained here, however, are insufficiently separated to generate the observed meson-mass splittings. The quartic term is necessary because it removes the oscillatory behavior produced by the quadratic term alone. It also decouples spontaneous from explicit chiral-symmetry breaking~\cite{Gherghetta:2009ac}. Nevertheless, the quartic term alone is insufficient to describe the meson multiplets. We therefore introduce an additional quadratic contribution in the form of a $z$-dependent mass term.

According to Ref.~\cite{Fang:2016nfj}, the $z$-dependent mass term produces a linear asymptotic behavior of $\chi_q$ at large $z$. A linear profile as $z\to\infty$ ensures that the mass difference between vector and axial-vector resonances approaches a constant, whereas $\chi_q(z\to\infty)\to\mathrm{constant}$ would imply chiral-symmetry restoration in the mass spectrum~\cite{Gherghetta:2009ac}. We therefore assume
\begin{equation}
    m_{5,q}^2(z) = -3 - \mu^2_q z^2,
\end{equation}
where $\mu_q$ is a flavor-dependent parameter to be determined.

In principle, $m_5^2(z)$ should be determined dynamically rather than introduced phenomenologically. According to the AdS/CFT dictionary, the UV behavior of $\chi_q(z)$ should be $\chi_q(z\to0)=m_q\zeta z+\frac{\sigma_q}{\zeta}z^3$. Substituting this expression into Eq.~\eqref{eq:EoM of chi} constrains the near-boundary dependence of $m_5^2(z)$:
\begin{equation}
    m_{5}^2(z \to 0)=-3-\mu^2_{q}(\mu_{c},m_{q},\kappa,\ldots)z^2+\cdots.
\end{equation}
Because $m_q$ appears in this expression, we allow $\mu_q$ to be flavor dependent. Rather than deriving the exact form of $m_5^2(z)$ or $\mu_q$, we adopt a phenomenological approach and determine $\mu_q$ by fitting the vector-meson spectrum. This term effectively accounts for the omitted backreaction of $\chi_q$ on the background fields and for corrections to the anomalous dimension~\cite{Alho:2013dka}.

\begin{figure}
    \centering
    \includegraphics[width=0.9\linewidth]{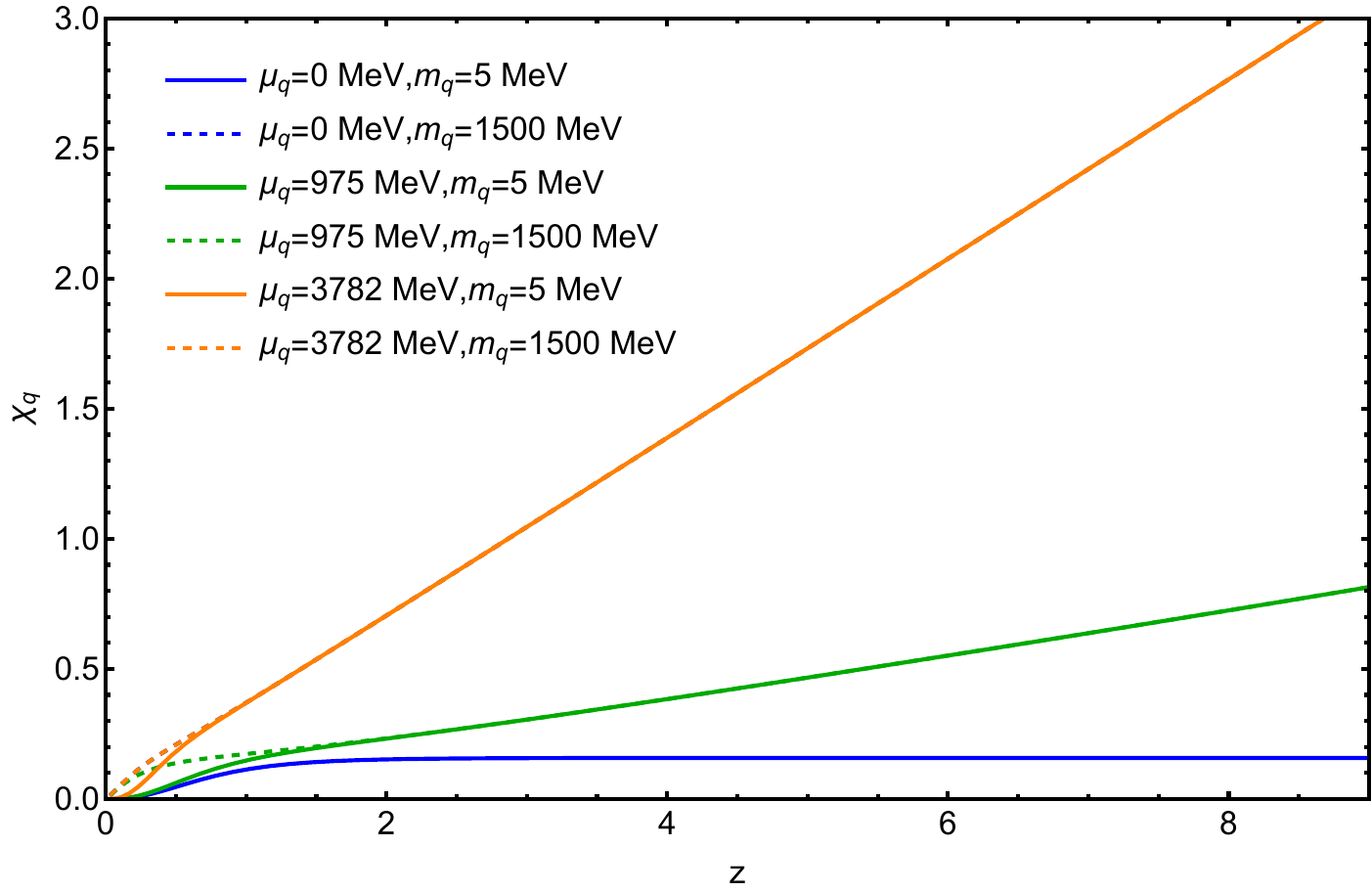}
    \caption{Scalar profiles $\chi_q$ obtained for different parameter values with $V(\chi_q)=(m_5^2-\mu_q^2z^2)\chi_q^2-\kappa\chi_q^4$. Unless otherwise indicated, $\mu_g^2=0.205\,\mathrm{GeV}^2$ and $T=60\,\mathrm{MeV}$ are used.}
    \label{fig:chi-z-modified}
\end{figure}

The scalar profiles obtained after including the five-dimensional mass correction are shown in Fig.~\ref{fig:chi-z-modified}. The flavor-dependent parameters $\mu_q$ separate the profiles for different flavors and produce a clear linear behavior at large $z$. Variations in $m_q$ affect mainly the UV region and have only a small influence on the observables considered here. These observables are controlled primarily by the infrared behavior of $\chi_q(z)$ and hence by $\mu_q$.

\subsubsection{Vector sector}
\label{sec:Vector sector}

We derive the equations of motion for the vector field from the second-order action $S_M^{(2)}$:
\begin{equation}
\begin{aligned}
    S_M^{(2)}={}&-\int d^5 x \sqrt{g}\,e^{-\Phi_m}
    \Bigg[ \frac{1}{4 g_5^2} F^{a,MN} F^a_{MN} \\
    &\qquad{}- M^{ab}_{V}V^{a,M}V^b_{M}\Bigg],
\end{aligned}
\end{equation}
where $F^{a,MN}=\partial^M V^{a,N}-\partial^N V^{a,M}$ and $M^{ab}_{V}=\mathrm{Tr}\{[t^a,X][t^b,X]\}$. Varying the action yields
\begin{equation}
\label{eq:EoM v}
    \partial_{M} (\sqrt{g}\,e^{-\Phi_m} F^{MN} ) + 2 g_5^2 \sqrt{g}\,e^{-\Phi_m} M^{aa}_{V}V^{a,N}=0.
\end{equation}
For $N_f=2$, $M_V^{ab}=0$, and the vector-field action is gauge invariant, allowing the choice $V_z=0$. For $N_f>2$, $M_V^{ab}\neq0$, and the broken gauge symmetry implies that the vector current is no longer conserved. However, $V_z$ and the four-dimensional longitudinal component have been shown to correspond to scalar mesons~\cite{Abidin:2009aj} and are therefore neglected here. Writing the remaining components explicitly gives
\begin{equation}
\sqrt{g} \, e^{-\Phi_m}\partial_{\mu}\partial^{z}V^{\mu}=0,
\end{equation}
\begin{equation}
\begin{aligned}
&V_{t}'' - \left( \frac{1}{z} + \Phi_m'-A_{s}' \right) V_{t}'
+ 2g_{5}^2 \frac{e^{2A_{s}}}{z^2f} M_{V} V_t \\
&\quad{}- \frac{e^{-B}}{f}\Big[ p_{x^1}^2 V_{t} +\omega \, p_{x^1} V_{x^1} \\
&\quad{}+ p_{x^2}^2 V_{t} +\omega \, p_{x^2} V_{x^2}
+ e^{3B}(p_{x^3}^2 V_{t} +\omega \, p_{x^3} V_{x^3}) \Big] =0,
\end{aligned}
\end{equation}
\begin{equation}
\begin{aligned}
&V_{x^1}'' - \left( \frac{1}{z} + B' + \Phi_m'-A_{s}' - \frac{f'}{f} \right) V_{x^1}' \\
&{}+ 2g_{5}^2 \frac{e^{2A_{s}}}{z^2f} M_{V} V_{x^1}
- \frac{1}{f}\Big[ e^{-B}(p_{x^2}^2 V_{x^1} - p_{x^1} \, p_{x^2} V_{x^2}) \\
&{}+ e^{2B}(p_{x^3}^2 V_{x^1} -p_{x^1} \, p_{x^3} V_{x^3})
-\frac{1}{f}(\omega^2 V_{x^1}+\omega\, p_{x^1}V_{t}) \Big] =0,
\end{aligned}
\end{equation}
\begin{equation}
\begin{aligned}
&V_{x^2}'' - \left( \frac{1}{z} + B' + \Phi_m'-A_{s}' - \frac{f'}{f} \right) V_{x^2}' \\
&{}+ 2g_{5}^2 \frac{e^{2A_{s}}}{z^2f} M_{V} V_{x^2}
- \frac{1}{f}\Big[ e^{-B}(p_{x^1}^2 V_{x^2} - p_{x^2} \, p_{x^1} V_{x^1}) \\
&{}+ e^{2B}(p_{x^3}^2 V_{x^2} -p_{x^2} \, p_{x^3} V_{x^3})
-\frac{1}{f}(\omega^2 V_{x^2}+\omega\, p_{x^2}V_{t}) \Big] =0,
\end{aligned}
\end{equation}
\begin{equation}
\begin{aligned}
&V_{x^3}'' - \left( \frac{1}{z} - 2B' + \Phi_m'-A_{s}' - \frac{f'}{f} \right) V_{x^3}' \\
&\quad{}+ 2g_{5}^2 \frac{e^{2A_{s}}}{z^2f} M_{V} V_{x^3}
- \frac{1}{f}\Big[ e^{-B}(p_{x^1}^2 V_{x^3} - p_{x^3} \, p_{x^1} V_{x^1} \\
&\quad{}+ p_{x^2}^2 V_{x^3} - p_{x^3} \, p_{x^2} V_{x^2} )
-\frac{1}{f}(\omega^2 V_{x^3}+\omega\, p_{x^3}V_{t}) \Big] =0.
\end{aligned}
\end{equation}
The last four equations are written in momentum space with all three spatial momentum components nonzero, rather than with one coordinate axis aligned with the momentum. The first equation imposes a constraint on the vector-field components, leaving three independent degrees of freedom corresponding to the three vector-meson polarizations. Because the $z$ component couples to the four-dimensional longitudinal component, neglecting this sector isolates the vector-meson degrees of freedom.

To isolate the independent degrees of freedom and simplify the coupled EoMs, we introduce the electric fields $E_i$:
\begin{equation}
    E_i(p,z) \equiv \omega V_i(p,z)+p_i V_t(p,z).
\end{equation}
Following the procedure of Ref.~\cite{Sheng:2024kgg}, we derive the equations of motion for $E_i(p,z)$ from Eq.~\eqref{eq:EoM v}.

For compactness in the derivation from Eq.~\eqref{eq:V0' to Vi'} to Eq.~\eqref{eq:E3 EoM}, we introduce the auxiliary metric \(h_{\mu\nu}=-f\,dt^2+e^B(dx_1^2+dx_2^2)+e^{-2B}dx_3^2\), so that \(p^2=h_{\mu\nu}p^\mu p^\nu\). The covariant momentum is \(p_\mu=(-\omega,p_1,p_2,p_3)\), and \(p^\mu=h^{\mu\nu}p_\nu\). Repeated spatial indices are summed.
From the definition of $E_i$ and the constraint equation, we obtain:
\begin{align}
V_{0}'&= - \frac{p^i}{p^0}V_{i}', \label{eq:V0' to Vi'} \\
E_{i}'&=-p_{0}\left( \delta_{i}^j + \frac{p_{i} p^j}{p^0 p_{0}} \right) V_{j}'.
\end{align}
Inverting the second equation expresses $V'_\mu$ in terms of $E'_i$:
\begin{align}
 V_i' &= -\frac{1}{p_0} \left( \delta_i^j - \frac{p_i p^j}{p^2} \right) E_j', \\
 V_0' &= \frac{p^i}{p_0 p^0} \left( \delta_i^j - \frac{p_i p^j}{p^2} \right) E_j'.
\end{align}
We then obtain the EoMs for $E_i$:
\begin{widetext}
\begin{align}
E_{\alpha}'' - \left( \frac{1}{z} + \Phi_m' -A_{s}' + B' -\frac{f'}{f} \right) E_{\alpha}'
+ 2g_{5}^2 \frac{e^{2A_{s}}}{z^2 f}M_{V}E_{\alpha}
{}- \frac{p^2}{f}E_{\alpha}
+ \left( B'- \frac{f'}{f} \right) \frac{p_{\alpha}}{p^2}p^iE_{i}' &= 0 , \\
E_{3}'' - \left( \frac{1}{z} + \Phi_m' -A_{s}' - 2B' -\frac{f'}{f} \right) E_{3}'
+ 2g_{5}^2 \frac{e^{2A_{s}}}{z^2 f}M_{V}E_{3}
{}- \frac{p^2}{f}E_{3}
- \left( 2B'+ \frac{f'}{f} \right) \frac{p_{3}}{p^2}p^iE_{i}' &=0 . \label{eq:E3 EoM}
\end{align}
\end{widetext}
where $\alpha=1,2$.

The matrix $M_V^{ab}$ is diagonal, with nonzero elements only for $a\neq1,2,3,8,15$. Consequently, the $\phi$, $J/\psi$, and $\rho$ mesons are degenerate within this model. Following Ref.~\cite{Chen:2021wzj}, we lift this degeneracy by introducing two auxiliary scalar fields, $H_s=\frac{1}{2}\text{diag}[0,0,h_s(z),0]$ and $H_c=\frac{1}{2}\text{diag}[0,0,0,h_c(z)]$, with the same five-dimensional mass and potential terms as $X$. We add the corresponding terms
\begin{equation}
    S=-\int{d^5x}\sqrt{g}e^{-\Phi_m}\sum_{q=s,c} \mathrm{Tr}\left[ \left| DH_q \right|^2+V(\left| H_q \right|) \right],
\end{equation}
where $D_{M}H_s=\partial_{M}H_s-i\{H_s,V^{8}_M\}$ and $D_{M}H_c=\partial_{M}H_c-i\{H_c,V^{15}_M\}$. A fully dynamical treatment would determine $h_q(z)$ by solving its EoM. As shown in the previous subsection, however, $\chi_q(z)$ is linear except in the UV region, where deviations from linearity have little impact on the observables considered here. We therefore approximate $h_s(z)=m_{hs}z$ and $h_c(z)=m_{hc}z$, with $m_{hs}$ and $m_{hc}$ treated as free parameters.
Accordingly, the modified vector-field EoMs are obtained by replacing $M_V^a$ in Eq.~\eqref{eq:EoM v} with $M_V^a-M_H^a$. Here $M_H^a$ is nonzero only for $a=8,15$, with $M_H^{8}=\frac{1}{3}h_s(z)^2$ and $M_H^{15}=\frac{3}{8}h_c(z)^2$.

\section{Spectral Functions and Mass Spectrum}
\label{sec:Spectral Function and Mass spectrum}

In this section, we calculate the retarded two-point correlation functions of the vector fields and extract the corresponding vector-meson spectral functions. We also determine the free parameters introduced in Sec.~\ref{sec:matter part} by fitting the vector-meson spectrum.

\subsection{The retarded correlator in holography}

We now solve the equations of motion and derive the retarded correlator.

First, we write down the on-shell action: 
\begin{equation}
S^{(2)}_{on-shell}= \left. \frac{1}{2g_5^2} \int d^4 x\, e^{-\Phi_m} \sqrt{g} \, g^{zz} g^{\mu\nu} V_\mu \partial_z V_\nu \right|_{z=0}.
\end{equation}
This action can be rearranged as $S_{on-shell}= \int \frac{d^4p}{(2\pi)^4} V_{0}^{\mu}(p) \mathcal{F}_{\mu \nu}(p)V_{0}^{\nu}(-p)$, where $V^{\mu}_0(p)\equiv V^{\mu}(p,z=0)$ is the boundary source of the vector field. Near the horizon $z=z_h$, a solution $V^{\mu}(p,z)$ of the EoMs has the asymptotic form
\begin{equation}
\label{eq:V/E IR expansion}
\begin{aligned}
    V_{\mu}(z)={}&(z_h - z)^{\pm \frac{i\,\omega}{4\pi T}}
    \Big[a_{\mu}^{(0)}+a_{\mu}^{(1)}\,(z_h - z) \\
    &\qquad{}+\mathcal{O}\left(z_h - z\right)^2\Big].
\end{aligned}
\end{equation}
The near-horizon expansion has two branches, corresponding to the incoming-wave condition (negative sign in the exponent) and the outgoing-wave condition (positive sign). The electric field $E_i(p,z)$ has the same near-horizon behavior. Selecting the incoming (outgoing) branch makes $\mathcal{F}_{\mu\nu}$ the retarded (advanced) two-point correlation function~\cite{Son:2002sd}.

Because the equations are formulated in terms of the electric field $E_i$, we first express the on-shell action in terms of $E_i$ and then rewrite its boundary values in terms of $V_0^\mu(p)$. The resulting action is
\begin{widetext}
\begin{equation}
\label{eq:on-shell action}
\begin{aligned}
   S^{(2)}_{on-shell} ={}& \frac{1}{2g_5^2} \frac{e^{A_{s}-\Phi_m}}{z}
   \int \frac{d^4 p}{(2\pi)^4} \frac{1}{p_{0}^2}
   \left( \frac{p^{i} p^j}{p^2}-\eta^{ij} \right)
   \mathcal{E}_{jm}(-p,z) \mathcal{E}_{in}'(p,z) \\
   &\quad{}\times\Big[ -p_{n}p_{m}V_{0}(-p)V_{0}(p)
   +p_{m}p_{0}V_{0}(-p)V_{n}(p)
   +p_{n}p_{0}V_{m}(-p)V_{0}(p)
   -p_{0}^2V_{m}(-p)V_{n}(p) \Big]
\end{aligned}
\Bigg|_{z=0}
\end{equation}
\end{widetext}

where $\mathcal{E}_{ij}$ is the bulk-to-boundary propagator of the electric field, defined by $E_i(p,z)=E_j(p)\mathcal{E}_{ij}(p,z)$ with $\mathcal{E}_{ij}(p,0)=\delta_{ij}$. Its matrix structure reflects the coupling among the electric-field components in the linearized EoMs. We then extract $\mathcal{F}^{\mu\nu}$ from Eq.~\eqref{eq:on-shell action} and obtain the correlation functions
\begin{equation}
\begin{aligned}
G^{00}(p)={}&\frac{1}{g_5^2}\frac{e^{A_s-\Phi_m}}{z}\frac{1}{p_0^2}\\
&\times\left(\frac{p^i p^j}{p^2}-\eta^{ij}\right)
\mathcal{E}_{jm}(-p,z)\mathcal{E}_{in}'(p,z)p_m p_n\\
={}&\frac{1}{g_5^2}\frac{e^{A_s-\Phi_m}}{z}
\frac{p^i p_n}{p^2}\mathcal{E}_{in}'(p,0),
\end{aligned}
\end{equation}
\begin{equation}
\begin{aligned}
G^{mn}(p)={}&\frac{1}{g_5^2}\frac{e^{A_s-\Phi_m}}{z}\frac{1}{p_0^2}\\
&\times\left(\frac{p^i p^j}{p^2}-\eta^{ij}\right)
\mathcal{E}_{jm}(-p,z)\mathcal{E}_{in}'(p,z)p_0^2\\
={}&\frac{1}{g_5^2}\frac{e^{A_s-\Phi_m}}{z}
\left(\frac{p^i p^m}{p^2}-\eta^{im}\right)\mathcal{E}_{in}'(p,0)
\end{aligned}
\end{equation}
\begin{equation}
\begin{aligned}
G^{0n}(p)={}&-\frac{1}{g_5^2}\frac{e^{A_s-\Phi_m}}{z}\frac{1}{p_0^2}\\
&\times\left(\frac{p^i p^j}{p^2}-\eta^{ij}\right)
\mathcal{E}_{jm}(-p,z)\mathcal{E}_{in}'(p,z)p_0p_m\\
={}&-\frac{1}{g_5^2}\frac{e^{A_s-\Phi_m}}{z}
\frac{p^i p_0}{p^2}\mathcal{E}_{in}'(p,0).
\end{aligned}
\end{equation}
\begin{equation}
\begin{aligned}
G^{n0}(p)={}&-\frac{1}{g_5^2}\frac{e^{A_s-\Phi_m}}{z}\frac{1}{p_0^2}\\
&\times\left(\frac{p^i p^j}{p^2}-\eta^{ij}\right)
\mathcal{E}_{jn}(-p,z)\mathcal{E}_{im}'(p,z)p_0p_m\\
={}&-\frac{1}{g_5^2}\frac{e^{A_s-\Phi_m}}{z}
\frac{p_m}{p_0}\left(\frac{p^i p^n}{p^2}-\eta^{in}\right)
\mathcal{E}_{im}'(p,0).
\end{aligned}
\end{equation}

The retarded two-point correlator $G^{R}_{\mu\nu}(\omega,\mathbf{p})$ encodes the spectral information of the vector current in the boundary theory. An isolated pole at $\omega=\omega_R-i\omega_I$ corresponds to a well-defined quasiparticle whose real part $\omega_R$ determines its energy and whose imaginary part $\omega_I>0$ determines its thermal width. Extracting the meson spectrum therefore amounts to locating the poles of $G^{R}_{\mu\nu}$. We identify these poles through a quasinormal-mode (QNM) analysis.

\subsection{Quasinormal modes and meson spectrum}

A quasinormal mode is a normal mode of a field propagating in a dissipative geometry. In the holographic setting, the bulk black hole acts as the dissipative system: it absorbs flux reaching the horizon, so no energy is reflected toward the boundary. A QNM is therefore defined as a solution of the bulk equation of motion that satisfies two boundary conditions simultaneously: an \emph{incoming-wave} condition at the horizon $z=z_h$, requiring the mode to propagate into the black hole, and a \emph{Dirichlet} condition at the AdS boundary $z=0$. The first is the condition imposed in Eq.~\eqref{eq:V/E IR expansion} to select the retarded correlator, while the second turns the problem into an eigenvalue problem. The complex frequencies $\omega_n$ that satisfy both conditions are the \emph{quasinormal frequencies}; they coincide with the poles of the retarded Green's function $G^{R}_{\mu\nu}$.

The link between QNMs and correlator singularities follows directly from the structure of the bulk-to-boundary propagator $\mathcal{E}_{ij}(p,z)$. Starting from the horizon, we solve the equations using the three linearly independent initial conditions $(a_{1}^{(0)},a_{2}^{(0)},a_{3}^{(0)})=(1,0,0)$, $(0,1,0)$, and $(0,0,1)$, and assemble the resulting solutions into the matrix
\begin{equation}
    \mathbb{E}(z)=
    \begin{pmatrix}
        E_{1}^{A}(z) & E_{1}^{B}(z) & E_{1}^{C}(z)\\[2pt]
        E_{2}^{A}(z) & E_{2}^{B}(z) & E_{2}^{C}(z)\\[2pt]
        E_{3}^{A}(z) & E_{3}^{B}(z) & E_{3}^{C}(z)
    \end{pmatrix}.
\end{equation}
A general incoming solution is a linear combination $E_{i}(z)=\mathbb{E}_{ij}(z)\,c_{j}$, and enforcing the boundary value $E_{i}(0)=E_{i}^{0}$ fixes $c_{j}=[\mathbb{E}(0)^{-1}]_{jk}E_{k}^{0}$. Comparison with $E_{i}(p,z)=\mathcal{E}_{ik}(p,z)\,E_{k}^{0}$ then gives
\begin{equation}
\label{eq:propagator}
    \mathcal{E}_{ik}(p,z)=\mathbb{E}_{ij}(z)\,[\mathbb{E}(0)^{-1}]_{jk}.
\end{equation}
Because $\mathbb{E}^{-1}=\mathrm{adj}\,\mathbb{E}/\det\mathbb{E}$, the correlators, which are proportional to $\mathcal{E}'(p,0)$, diverge whenever
\begin{equation}
\label{eq:QNM condition}
    \det\mathbb{E}(\omega,\mathbf{p},z=0)=0.
\end{equation}
Eq.~\eqref{eq:QNM condition} is the QNM condition in the general coupled case. In special kinematic configurations in which the $E_i$ decouple, for instance when the spatial momentum is aligned with a symmetry axis, the propagator becomes diagonal and the condition simplifies to $E_i(\omega,\mathbf{p},0)=0$. We note, however, that $\det\mathbb{E}=0$ is only a necessary condition: a candidate frequency is an actual singularity of a given correlator component only if the corresponding projected propagator has a nonzero residue.

We now apply this method to determine the vacuum vector-meson masses, which fix the free parameters of the matter sector introduced in Sec.~\ref{sec:matter part}. At finite temperature, the notion of particle mass is not unique because the heat bath breaks Lorentz invariance and distinguishes the temporal and spatial directions. Two commonly used definitions are the \emph{pole mass} and the \emph{screening mass}, which characterize two-point correlation functions in the temporal and spatial directions, respectively~\cite{Cao:2021tcr}. Here we focus on the pole mass, which is extracted from the dispersion relation by locating the correlator pole at vanishing spatial momentum:
\begin{equation}
    m_{\rm pole} = \omega_{R}(\mathbf{p}=0).
\end{equation}
It therefore characterizes the energy of a quasiparticle at rest.

In the confined, low-temperature regime, the thermal medium has only a mild effect on the dispersion relation, and the pole masses differ only slightly from their $T=0$ values. A sufficiently low-temperature pole mass therefore provides a reliable approximation to the vacuum meson mass. Numerically, a direct $T=0$ calculation also requires truncating the holographic coordinate $z$, introducing errors comparable to those of the low-temperature approximation.

We therefore set $\mathbf{p}=0$, solve the QNM condition~\eqref{eq:QNM condition} at low temperature ($z_h=20\,\mathrm{GeV}^{-1}$ and $T=28.1\,\mathrm{MeV}$) and zero chemical potential, and identify the real part of the QNM frequency as the vacuum mass of the corresponding vector meson:
\begin{equation}
\label{eq:vacuum mass}
    m_{\rm vacuum}\;\simeq\;\omega_{R}(\mathbf{p}=0,\,T\ll T_{c}).
\end{equation}
We next fit the model results to experimental data~\cite{ParticleDataGroup:2024cfk}. We determine $\mu_g$ from the $\rho$-meson mass and determine $\kappa$ and $\mu_l$ from the masses of the $a_1$ meson and its radial excitations. Although the model does not explicitly include the axial-vector field, the corresponding calculation is obtained by replacing the commutator in $M_V$ with an anticommutator and multiplying the result by $-1$~\cite{Abidin:2009aj}. We then determine $\mu_s$, $\mu_c$, $h_s$, and $h_c$ from the masses of the $K^*$, $D^*$, $\phi$, and $J/\psi$ mesons, respectively. Because the observables considered here are only weakly sensitive to the quark masses, we set $m_l=5\,\mathrm{MeV}$, $m_s=93\,\mathrm{MeV}$, and $m_c=1273\,\mathrm{MeV}$. The resulting meson spectrum and parameter values are listed in Tab.~\ref{tab:meson spectrum}.

\begin{table*}[tbp]
    \centering
    \begin{ruledtabular}
    \begin{tabular}{cccc@{\hspace{3em}}cccc}
        Meson & $n$ & Model (MeV) & Experiment (MeV) & Meson & $n$ & Model (MeV) & Experiment (MeV) \\ \hline
        $\rho$ & 0 & 773 & $775.26\pm0.23$ & $\phi$ & 0 & 1016 & $1019.460\pm0.016$ \\
        & 1 & 1206 & $1282\pm37$ & $D^{*}$ & 0 & 1999 & $2006.85\pm0.05$ \\
        & 2 & 1510 & $1465\pm25$ & & 1 & 2751 & $2627\pm10$ \\
        & 3 & 1759 & $1720\pm20$ & $D^{*}_{s}$ & 0 & 1807 & $2106.6\pm3.4$ \\
        $a_{1}$ & 0 & 1342 & $1230\pm40$ & & 1 & 2491 & $2714\pm5$ \\
        & 1 & 1622 & $1655\pm16$ & $J/\psi$ & 0 & 3087 & $3096.900\pm0.009$ \\
        & 2 & 1857 & $1930^{+30}_{-70}$ & \multicolumn{4}{c}{\textbf{Fitted parameters}} \\ \cline{5-8}
        & 3 & 2066 & $2096\pm122$ & \multicolumn{2}{c}{$\mu_g = 451$ MeV} & \multicolumn{2}{c}{$\kappa = -240$} \\
        $K^{*}$ & 0 & 892 & $891.67\pm0.26$ & \multicolumn{2}{c}{$\mu_l = 316$ MeV} & \multicolumn{2}{c}{$\mu_s = 1068$ MeV} \\
        & 1 & 1361 & $1414\pm15$ & \multicolumn{2}{c}{$\mu_c = 3834$ MeV} & \multicolumn{2}{c}{$h_s = 47$ MeV} \\
        & 2 & 1683 & $1718\pm18$ & \multicolumn{2}{c}{$h_c = 300$ MeV} & &
    \end{tabular}
    \end{ruledtabular}
    \caption{\label{tab:meson spectrum}Vector meson spectrum: model results versus experimental values \cite{ParticleDataGroup:2024cfk}. $n=0$ denotes the ground state and $n=1,2,3$ label radial excitations. The fitted parameters of the matter action are listed in the lower-right block.}
\end{table*}

The relatively low $D_s^*$ mass predicted by the model deserves further comment. In the present matter-sector construction, the flavor-dependent mass terms in the EoMs of the open-charm vector mesons are controlled by differences between the corresponding scalar profiles:
\begin{align*}
M_V^{D^*}(z)&\propto [\chi_c(z)-\chi_l(z)]^2,\\
M_V^{D_s^*}(z)&\propto [\chi_c(z)-\chi_s(z)]^2,
\end{align*}
up to a common normalization. For the fitted scalar profiles, $\chi_s(z)$ lies closer to $\chi_c(z)$ than $\chi_l(z)$ does over the relevant bulk region, leading to $m_{D_s^*}<m_{D^*}$ within the present ansatz. This ordering is a structural consequence of the adopted flavor construction rather than a numerical uncertainty. The discrepancy between the predicted and experimental $D_s^*$ masses is therefore a limitation of the current matter sector. Nevertheless, within this parametrization, the qualitative trend of the $D_s^*$ spin alignment remains unchanged, as shown in the next section.

\subsection{Spectral Functions}
\label{sec:spectral function}

The spectral function is defined as minus the imaginary part of the retarded two-point correlation function in momentum space:
\begin{equation}
\label{eq:spectral fun}
    \varrho_{\alpha\beta}(p) \equiv -\operatorname{Im} G^{R}_{\alpha\beta}(p).
\end{equation}
Since our ultimate goal is the spin alignment of vector mesons, the quantities of direct interest are not the individual tensor components $\varrho_{\alpha\beta}$ themselves, but the spectral functions projected onto definite spin-polarization states.

To separate the contributions of the different spin states, we decompose the spectral function in a complete basis of polarization vectors:
\begin{equation}
    \varrho^{\mu\nu}(p)=\sum_{\lambda,\lambda'=0,\pm1}v^\mu(\lambda,p)\,v^{*\nu}(\lambda',p)\,\tilde{\varrho}_{\lambda\lambda'}(p),
    \label{eq:rho decomposition}
\end{equation}
where the polarization vectors are
\begin{equation}
    v^\mu(\lambda,p)=\left(\frac{\mathbf{p}\cdot\boldsymbol{\epsilon}_\lambda}{M},\;\boldsymbol{\epsilon}_\lambda+\frac{\mathbf{p}\cdot\boldsymbol{\epsilon}_\lambda}{M(\omega+M)}\,\mathbf{p}\right),
    \label{eq:polarization vector}
\end{equation}
where $M\equiv\sqrt{\omega^{2}-\mathbf{p}^{2}}$ denotes the invariant mass. The polarization vectors satisfy the orthonormality and completeness conditions
\begin{equation}
\begin{aligned}
    \eta_{\mu\nu}v^\mu(\lambda,p)v^{*\nu}(\lambda',p)
    &=\delta_{\lambda\lambda'},\\
    \sum_\lambda v^\mu(\lambda,p)v^{*\nu}(\lambda,p)
    &=\eta^{\mu\nu}+\frac{p^\mu p^\nu}{p^2}.
\end{aligned}
    \label{eq:ortho complete}
\end{equation}
The three-component vectors $\boldsymbol{\epsilon}_\lambda$ specify the spin directions in the meson rest frame, with $\boldsymbol{\epsilon}_0$ along the quantization axis and $\boldsymbol{\epsilon}_{\pm1}$ orthogonal to it. In heavy-ion collisions, spin alignment is measured relative to the global orbital angular momentum. We therefore choose the quantization axis along the rotation axis of the medium, i.e., the symmetry axis $x_3$ of the anisotropic metric in Eq.~\eqref{eq:metric}:
\begin{equation}
    \boldsymbol{\epsilon}_0=(0,0,1),\qquad
    \boldsymbol{\epsilon}_{\pm1}=\mp\frac{1}{\sqrt{2}}\,(1,\pm i,0).
    \label{eq:epsilon basis}
\end{equation}
Projecting the holographic correlator onto these vectors gives the spectral function in spin space,
\begin{equation}
    \tilde{\varrho}_{\lambda\lambda'}(p)=v_\mu(\lambda,p)\,v^*_\nu(\lambda',p)\,\varrho^{\mu\nu}(p).
    \label{eq:rho spin space}
\end{equation}
Because the two-point correlation tensor is symmetric in the present model, Eq.~\eqref{eq:rho spin space} gives $\tilde{\varrho}_{11}=\tilde{\varrho}_{-1-1}$.

\begin{figure}
    \centering
    \includegraphics[width=1\linewidth]{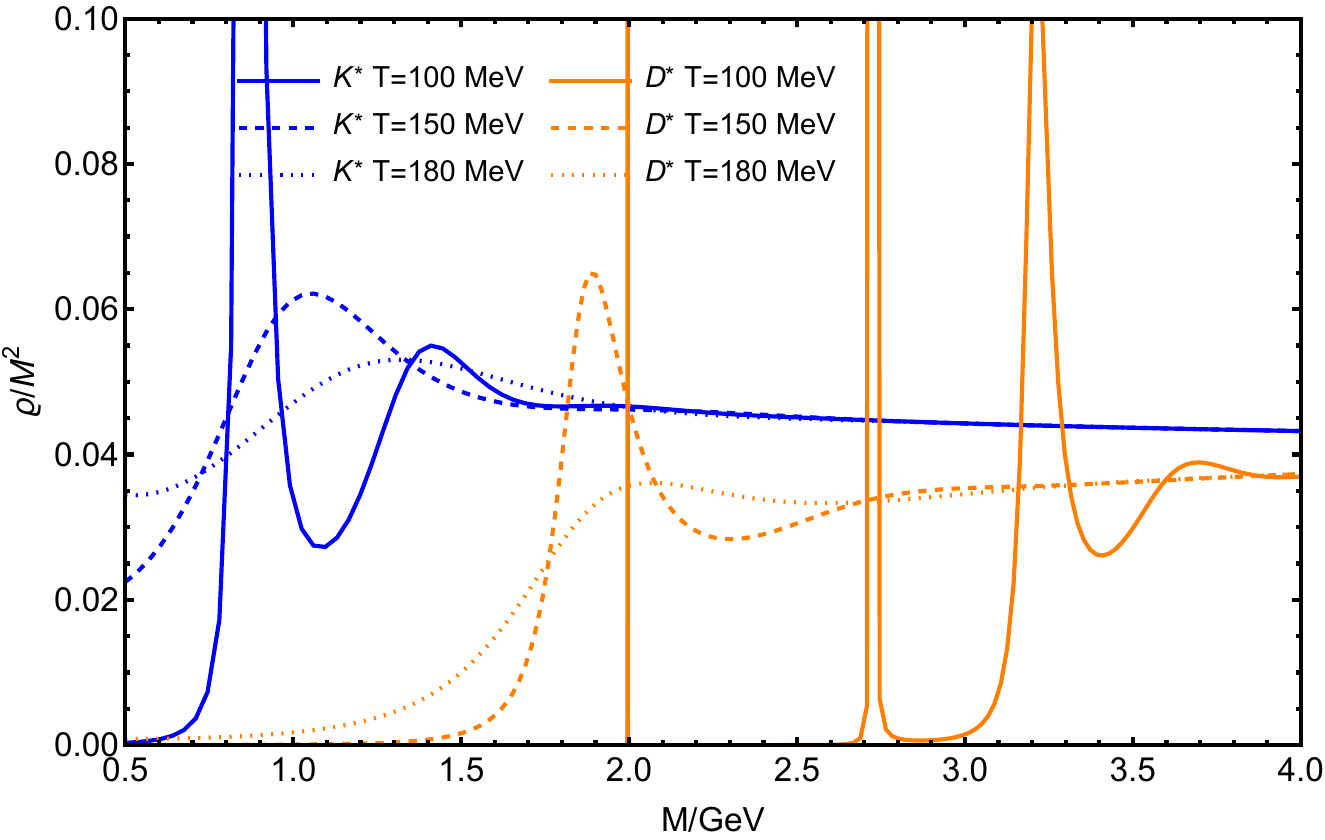}
    \caption{Spectral functions of the $K^{*}$ (blue) and $D^{*}$ (orange) mesons at zero spatial momentum and vanishing angular velocity, where the three polarization states are degenerate, plotted as $\varrho/M^{2}$ versus the invariant mass $M=\omega$. Solid, dashed, and dotted lines correspond to $T=100$, $150$, and $180\,\mathrm{MeV}$, respectively.}
    \label{fig:Spectral_function}
\end{figure}

Fig.~\ref{fig:Spectral_function} shows the spectral functions of the $K^{*}$ and $D^{*}$ mesons at zero spatial momentum and vanishing angular velocity. In this limit, the three polarization states are degenerate, and $\varrho$ denotes any diagonal component $\tilde{\varrho}_{\lambda\lambda}$. Because $\mathbf{p}=0$, the frequency equals the invariant mass, $\omega=M$, and we plot the rescaled quantity $\varrho/M^2$ at three temperatures. At low temperature, the spectral function consists of a tower of peaks at the real parts of the QNM frequencies. Their positions closely match the vacuum meson masses listed in Tab.~\ref{tab:meson spectrum}, while their widths reflect the lifetimes of the corresponding quasiparticles. As the temperature increases, the QNM poles move away from the real axis: each peak broadens and shifts away from its vacuum position, while radial excitations are washed out earlier than the ground state. The thermal widths of the heavier mesons grow more slowly with temperature.

\begin{figure}
    \centering
    \includegraphics[width=1\linewidth]{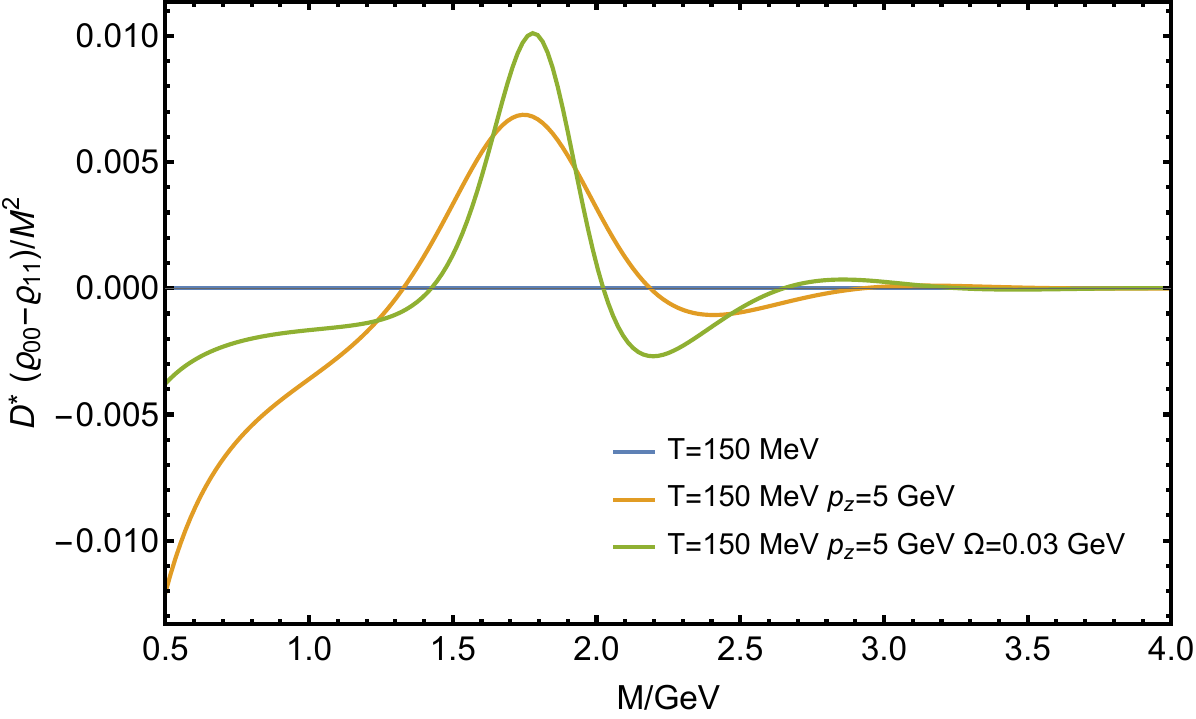}
    \caption{Difference between the longitudinal and transverse spin components of the $D^{*}$ spectral function, $(\tilde{\varrho}_{00}-\tilde{\varrho}_{11})/M^{2}$, at $T=150\,\mathrm{MeV}$, with the quantization axis along the rotation axis $x_3$. The difference vanishes for a meson at rest in a non-rotating medium (blue). A momentum $p_{z}=5\,\mathrm{GeV}$ along the rotation axis splits the spectral functions of the different spin states (orange), and an additional angular velocity $\Omega=0.03\,\mathrm{GeV}$ further enhances the splitting (green).}
    \label{fig:Ds_rho_compare}
\end{figure}

The spin dependence that is absent from Fig.~\ref{fig:Spectral_function} is shown in Fig.~\ref{fig:Ds_rho_compare}, which displays $(\tilde{\varrho}_{00}-\tilde{\varrho}_{11})/M^{2}$ for the $D^{*}$ meson at $T=150\,\mathrm{MeV}$. When the meson moves through the medium with momentum $p_z=5\,\mathrm{GeV}$ along the rotation axis, its momentum selects a preferred direction. The longitudinal and transverse channels then obey different EoMs, and their spectral functions split. Their difference develops a pronounced positive peak near the in-medium ground-state mass shell, indicating small differences among the positions, heights, and widths of the quasiparticle peaks for the $\lambda=0$ and $\lambda=\pm1$ states. Rotation further modifies this splitting: at $\Omega=0.03\,\mathrm{GeV}$, the peak in the difference is visibly enhanced.

This spin dependence of the spectral function is the origin of spin alignment in the present holographic framework. As shown in Sec.~\ref{sec:Spin Alignment}, any difference between $\tilde{\varrho}_{00}$ and $\tilde{\varrho}_{11}$ near the vacuum mass shell translates directly into a deviation of $\rho_{00}$ from the unpolarized value $1/3$. The sign and magnitude of this deviation are governed by how the hot and rotating medium reshapes the quasiparticle peaks of the different polarization states.

\section{Spin Alignment of Vector Mesons}
\label{sec:Spin Alignment}

The spin state of a vector meson is described by a $3\times3$ spin-density matrix $\rho_{\lambda\lambda'}$, where $\lambda,\lambda'=0,\pm1$ denote spin projections along a chosen quantization axis. Its $00$ element, $\rho_{00}$, is the probability of finding the meson in the $\lambda=0$ state and is conventionally called the spin alignment. It can be extracted from the polar-angle distribution of one decay daughter in the meson rest frame, measured relative to the quantization axis and integrated over the azimuthal angle. For $P$-wave strong decays into two pseudoscalar mesons, such as $\phi\to K^+K^-$ or $K^{*0}\to K\pi$, the meson spin is encoded in the orbital angular momentum of the daughters, and the distribution is
\begin{equation}
    W(\theta)=\frac{3}{4}\left[(1-\rho_{00})+(3\rho_{00}-1)\cos^2\theta\right].
    \label{eq:Wtheta strong}
\end{equation}
For dilepton decays such as $(J/\psi,\phi)\to\ell^+\ell^-$, where the spin is carried by the lepton pair, the distribution is usually written in terms of an anisotropy parameter $\lambda_\theta$:
\begin{equation}
    W(\theta)\propto\frac{1}{3+\lambda_\theta}\left(1+\lambda_\theta\cos^2\theta\right),
    \label{eq:Wtheta}
\end{equation}
where $\lambda_\theta$ is related to the spin alignment by
\begin{equation}
    \lambda_\theta=\frac{1-3\rho_{00}}{1+\rho_{00}}.
    \label{eq:lambda theta}
\end{equation}
In both channels, an isotropic distribution is recovered for $\rho_{00}=1/3$ ($\lambda_\theta=0$), corresponding to an unpolarized ensemble. The limiting cases $\rho_{00}=1$ ($\lambda_\theta=-1$) and $\rho_{00}=0$ ($\lambda_\theta=+1$) correspond to purely longitudinal and purely transverse polarization with respect to the quantization axis, respectively.

We now use the in-medium spectral functions obtained in Sec.~\ref{sec:spectral function} to calculate vector-meson spin alignment. Following Ref.~\cite{Sheng:2024kgg}, we use the dilepton production of flavorless vector mesons as an illustrative example.

For a lepton pair with total momentum $p^\mu=(\omega,\mathbf{p})$, the differential dilepton production rate $n(x,p)\equiv dN/(d^4x\,d^4p)$ is~\cite{Gale:1990pn}
\begin{multline}
    n(x,p)=-\frac{2g_{M\ell\bar\ell}^2}{3(2\pi)^5}\left(1-\frac{2m_\ell^2}{p^2}\right)\sqrt{1+\frac{4m_\ell^2}{p^2}}\;p^2\,n_B(x,\omega)\\
    \times\left(\eta_{\mu\nu}+\frac{p_\mu p_\nu}{p^2}\right)G_A^{\mu\alpha}(p)\,\varrho_{\alpha\beta}(x,p)\,G_R^{\beta\nu}(p),
    \label{eq:production rate n}
\end{multline}
where $m_\ell$ is the lepton mass, $g_{M\ell\bar\ell}$ is the meson--dilepton coupling strength, $n_B(x,\omega)=1/(e^{\omega/T(x)}-1)$ is the Bose--Einstein distribution, and $\eta_{\mu\nu}=\mathrm{diag}(-1,1,1,1)$ is the flat metric. The quantity $\varrho_{\alpha\beta}(x,p)$ is the thermal vector-current spectral function, defined as minus the imaginary part of the retarded current-current correlator computed holographically in Sec.~\ref{sec:spectral function}; its $x$ dependence enters through the local temperature $T(x)$. In contrast, $G_{R/A}^{\mu\nu}$ are vacuum propagators describing the freely propagating meson after freeze-out:
\begin{equation}
    G_{R/A}^{\mu\nu}(p)=-\frac{\eta^{\mu\nu}+p^\mu p^\nu/p^2}{p^2+m_V^2\pm i\,m_V\Gamma},
    \label{eq:vacuum prop}
\end{equation}
where $m_V$ and $\Gamma$ are the vacuum mass and width of the meson, respectively. In Eq.~\eqref{eq:production rate n}, the in-medium spectral function is sandwiched between vacuum propagators, thereby implementing an \emph{instantaneous freeze-out} assumption under which the detailed kinematics of the freeze-out stage are neglected. The spectral function $\varrho_{\alpha\beta}$ characterizes the in-medium quark--antiquark pair that is converted into a freely propagating meson at freeze-out. Consequently, Eq.~\eqref{eq:production rate n} has a Breit--Wigner form peaked at $-p^2=m_V^2$, so configurations with $M\approx m_V$ dominate the final dilepton yield.

To separate the contributions of the different spin states, we decompose the spectral function in the polarization-vector basis introduced in Sec.~\ref{sec:spectral function}, with the spin-space components $\tilde{\varrho}_{\lambda\lambda'}$ given by Eq.~\eqref{eq:rho spin space}. In a locally equilibrated medium, the decomposition in Eq.~\eqref{eq:rho decomposition} applies point by point, so $\varrho^{\mu\nu}(x,p)$ and $\tilde{\varrho}_{\lambda\lambda'}(x,p)$ acquire their $x$ dependence through the local thermodynamic parameters. Substituting this decomposition into Eq.~\eqref{eq:production rate n} separates the production rate into contributions from the individual spin states:
\begin{equation}
\begin{aligned}
    n_\lambda(x,p)={}&-\frac{2g_{M\ell\bar\ell}^2}{3(2\pi)^5}
    \left(1-\frac{2m_\ell^2}{p^2}\right)
    \sqrt{1+\frac{4m_\ell^2}{p^2}} \\
    &\qquad{}\times
    \frac{p^2\,n_B(x,\omega)\,\tilde{\varrho}_{\lambda\lambda}(x,p)}
    {(p^2+m_V^2)^2+m_V^2\Gamma^2},
\end{aligned}
    \label{eq:n lambda}
\end{equation}
with the total yield $n=\sum_{\lambda=0,\pm1}n_\lambda$. The spin alignment is then defined as the probability of the spin-zero state,
\begin{equation}
    \rho_{00}(x,\mathbf{p})\equiv\frac{\int d\omega\,n_0(x,p)}{\sum_{\lambda=0,\pm1}\int d\omega\,n_\lambda(x,p)}.
    \label{eq:rho00 def}
\end{equation}
Because each $n_\lambda$ is nonnegative, $0\leq\rho_{00}\leq1$ by construction.

The derivation above applies to the dilepton channel, which is available only for flavorless mesons. For open-flavor mesons, spin alignment is instead measured through strong decays, such as $K^{*}\to K\pi$. Nevertheless, the different decay channels can be treated analogously, and their differential production rates can be written in the unified form
\begin{equation}
n_\lambda(x,p)=g(p^2)\frac{n_B(x,\omega)\,\tilde{\varrho}_{\lambda\lambda}(x,p)}{(p^2+m_V^2)^2+m_V^2\Gamma^2},
    \label{eq:general n lambda}
\end{equation}
where the dynamical factor $g(p^2)$, arising from the interaction vertex of the specific decay channel, carries all the channel dependence. Moreover, when the width $\Gamma$ is much smaller than $m_V$, the denominator of Eq.~\eqref{eq:general n lambda} acts as a narrow peak around $-p^2=m_V^2$, within which $g(p^2)$ is common to all spin states and cancels in the ratio, so that the spin alignment is well approximated by the ratio of spectral functions evaluated on the mass shell,
\begin{equation}
    \rho_{00}\simeq\frac{\tilde{\varrho}_{00}}{\sum_{\lambda}\tilde{\varrho}_{\lambda\lambda}}\bigg|_{\omega=\sqrt{m_V^2+\mathbf{p}^2}}.
    \label{eq:rho00 narrow}
\end{equation}
This relation shows explicitly that, in the narrow-width limit, spin alignment is independent of the decay channel because the spectral functions characterize the medium rather than the decay process. A nonzero deviation of $\rho_{00}$ from $1/3$ arises whenever the spectral function of the longitudinal state ($\lambda=0$) differs from those of the transverse states ($\lambda=\pm1$), i.e., whenever rotational symmetry among the spin states is broken by either the motion of the meson relative to the medium or the intrinsic anisotropy of the medium.

As a numerical check of the narrow-width approximation, we evaluate the energy-integrated expression in Eq.~\eqref{eq:general n lambda} for the $K^*$ meson, which has the largest vacuum width among the mesons considered here. We use the dilepton form of $g(p^2)$ in Eq.~\eqref{eq:n lambda} as a representative smooth dynamical factor; alternative choices have only a minor effect. For $\Gamma_{K^*}\simeq50\,\mathrm{MeV}$, the difference between the full energy-integrated result and the on-shell approximation in Eq.~\eqref{eq:rho00 narrow} is of order $10^{-3}$. This finite-width correction is smaller than the corresponding peak-to-peak angular variation of $\rho_{00}$, which is of order $10^{-2}$. The approximation therefore preserves the qualitative pattern discussed below. Because the remaining mesons have substantially smaller vacuum widths, their freeze-out kernels are more strongly localized around the mass shell, and Eq.~\eqref{eq:rho00 narrow} is expected to be even more accurate for them.

\begin{figure*}[!t]
    \centering
    \includegraphics[width=0.96\textwidth]{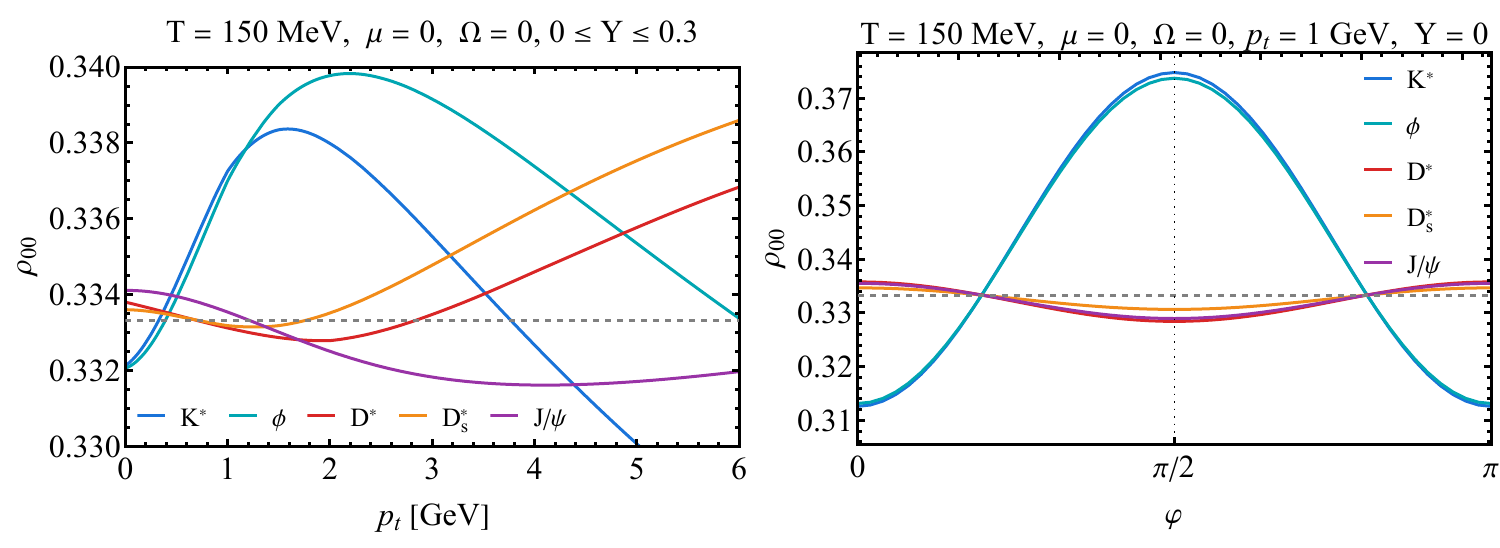}
    \caption{Overview of the spin alignment of the five vector mesons. Left panel: $\bar\rho_{00}(p_t)$ at $T=150\,\mathrm{MeV}$ and $\mu=\Omega=0$, additionally averaged over $0\leq Y\leq0.3$. Right panel: angular distributions at $Y=0$ and $p_t=1\,\mathrm{GeV}$, with the remaining parameters unchanged. The horizontal gray line denotes the unpolarized value $1/3$.}
    \label{fig:rho00 overview}
    \includegraphics[width=0.96\textwidth]{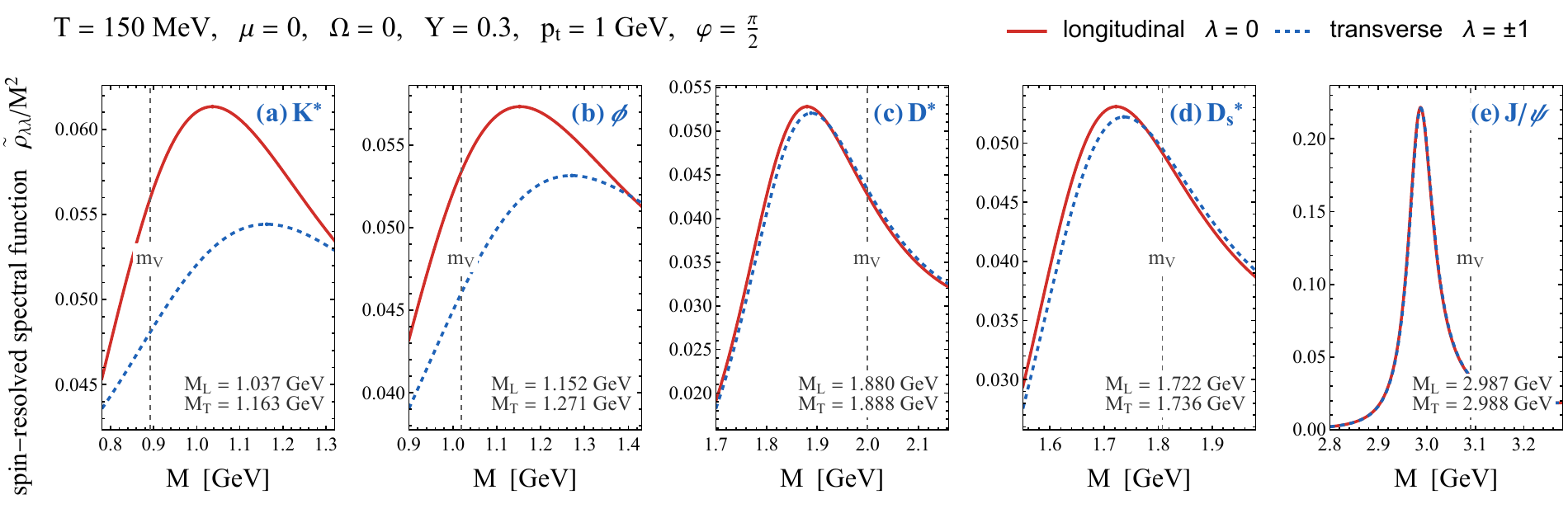}
    \caption{Longitudinal (red solid) and transverse (blue dashed) spin-resolved spectral functions of the five mesons at $T=150\,\mathrm{MeV}$, $\mu=\Omega=0$, $Y=0.3$, $p_t=1\,\mathrm{GeV}$, and $\varphi=\pi/2$. The vertical line denotes the vacuum mass $m_V$; $M_L$ and $M_T$ denote the longitudinal and transverse peak positions, respectively. }
    \label{fig:five meson polarized spectra}
\end{figure*}

\subsection{Results}
\label{sec:result}

We present results for the $K^*$, $\phi$, $D^*$, $D_s^*$, and $J/\psi$ mesons. The spin-quantization basis $\boldsymbol{\epsilon}_\lambda$ is the global event-plane basis specified in Sec.~\ref{sec:spectral function}, with the quantization axis along the rotation axis $x_3$ of the anisotropic metric in Eq.~\eqref{eq:metric}. To facilitate comparison with experiment, we parametrize the momentum in terms of transverse momentum $p_t$, azimuthal angle $\varphi$, and rapidity $Y$:
\begin{equation}
\begin{aligned}
    p^\mu&=\left(m_T\cosh Y,\;m_T\sinh Y,\;p_t\cos\varphi,\;p_t\sin\varphi\right),\\
    m_T&=\sqrt{m_V^2+p_t^2}.
\end{aligned}
    \label{eq:momentum param}
\end{equation}
Unless stated otherwise, $\bar\rho_{00}$ denotes an azimuthal average with the elliptic-flow weight adopted in Ref.~\cite{Sheng:2024kgg}:
\begin{equation}
\begin{aligned}
    \bar\rho_{00}(p_t,Y)&=
    \frac{\int_0^{2\pi}d\varphi\,[1+2v_2\cos(2\varphi)]\rho_{00}(p_t,Y,\varphi)}
    {\int_0^{2\pi}d\varphi\,[1+2v_2\cos(2\varphi)]},\\
    v_2&=0.15.
\end{aligned}
    \label{eq:azimuthal average}
\end{equation}

\subsubsection{Overview and spectral origin of the two patterns}

Fig.~\ref{fig:rho00 overview} summarizes the central result. At low transverse momentum, the five mesons separate into two classes: $K^*$ and $\phi$ have $\bar\rho_{00}<1/3$, whereas $D^*$, $D_s^*$, and $J/\psi$ have $\bar\rho_{00}>1/3$. As $p_t$ increases, the light-meson curves initially rise while the heavy-meson curves decrease, followed by species-dependent turning points and crossings. The opposite angular modulations are shown more clearly in the right panel. At $p_t=1\,\mathrm{GeV}$, the $K^*$ and $\phi$ curves reach a maximum when the momentum is aligned with the spin-quantization axis, whereas the three charmed-meson curves reach a minimum in the same direction.

The origin of this separation can be read directly from the polarized spectral functions. Fig.~\ref{fig:five meson polarized spectra} compares the longitudinal component $\tilde\varrho_{00}$ with the degenerate transverse components $\tilde\varrho_{11}=\tilde\varrho_{-1-1}$ at a representative common kinematic point. The vertical line marks the vacuum mass $m_V$, corresponding to the mass shell selected by the freeze-out kernel in Eq.~\eqref{eq:general n lambda}. For $K^*$ and $\phi$, the vacuum mass lies below both in-medium peaks, $m_V<M_L<M_T$. The longitudinal peak is therefore closer to the sampled mass shell, so $\tilde\varrho_{00}(m_V)>\tilde\varrho_{11}(m_V)$ and $\rho_{00}>1/3$ at $\varphi=\pi/2$. For $D^*$, $D_s^*$, and $J/\psi$, both peaks lie below the vacuum mass, $M_L<M_T<m_V$. On the high-mass flank, the transverse peak is closer to $m_V$, reversing the on-shell ordering and giving $\rho_{00}<1/3$. The splitting is particularly small for $J/\psi$, but the narrow spectral peak makes even a small displacement significant on the mass shell.

\begin{figure}[t]
    \centering
    \includegraphics[width=0.98\linewidth]{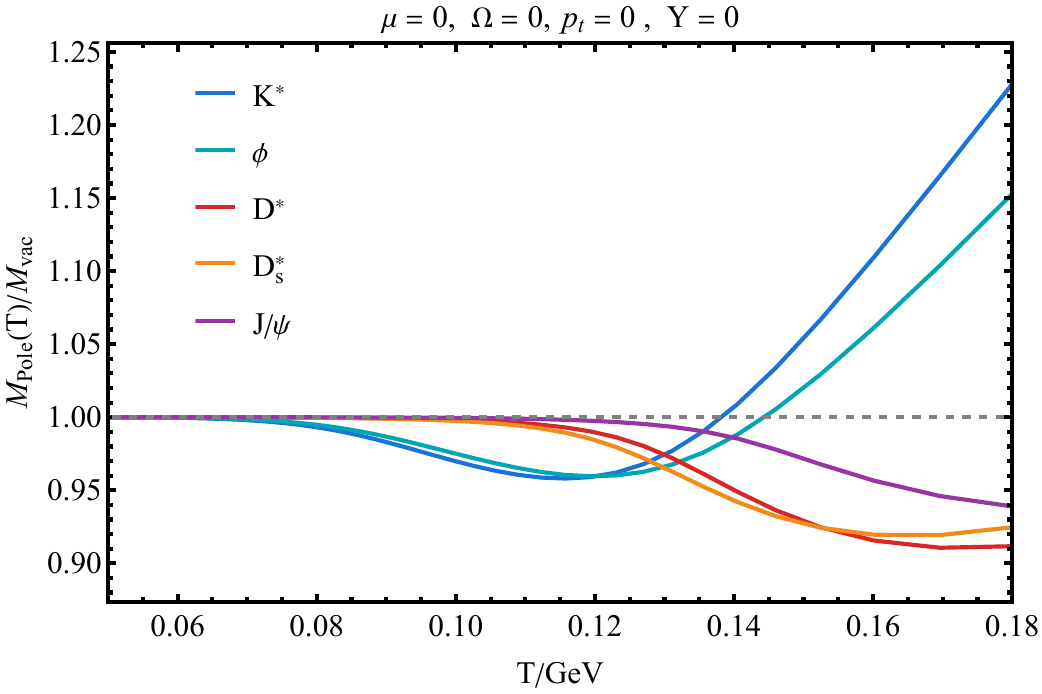}
    \caption{Temperature dependence of the pole masses normalized by their vacuum values at $\mu=\Omega=0$ and vanishing spatial momentum, where the polarization states are degenerate. The horizontal dashed line denotes $M_{\mathrm{pole}}(T)=M_{\mathrm{vac}}$. Around $T=150\,\mathrm{MeV}$, the $K^*$ and $\phi$ pole masses lie above their vacuum values, whereas the $D^*$, $D_s^*$, and $J/\psi$ pole masses lie below them.}
    \label{fig:pole mass overview}
\end{figure}

The opposite locations of the vacuum mass relative to the in-medium peaks can be traced to the different thermal pole-mass shifts of the two meson groups. At vanishing spatial momentum, the polarization states are degenerate and share a thermal pole mass $M_{\mathrm{pole}}(T)$. As shown in Fig.~\ref{fig:pole mass overview}, the pole masses of $K^*$ and $\phi$ shift upward relative to their vacuum values around $T=150\,\mathrm{MeV}$, whereas those of $D^*$, $D_s^*$, and $J/\psi$ shift downward. Consequently, the vacuum mass lies below the thermal peak for the first group and above it for the second. At finite momentum, this common thermal peak splits into longitudinal and transverse peaks $M_L$ and $M_T$ while retaining the relative ordering shown in Fig.~\ref{fig:five meson polarized spectra}.

Peak positions alone do not completely determine $\rho_{00}$. The relevant quantities are the full longitudinal and transverse spectral strengths near $m_V$, including differences in the peak positions, heights, widths, and background contributions. Nevertheless, Fig.~\ref{fig:pole mass overview} explains why the vacuum mass lies on opposite sides of the thermal peaks for the two meson groups, while Fig.~\ref{fig:five meson polarized spectra} shows how the subsequent polarization splitting produces opposite longitudinal--transverse spectral imbalances on the mass shell. Together, these results provide a unified explanation of the two qualitative spin-alignment patterns in terms of the mass-dependent thermal evolution and polarization splitting of the in-medium spectral functions.

\subsubsection{Dependence on kinematic and medium parameters}

The preceding subsection focused mainly on the low-$p_t$ and low-rapidity regions at fixed temperature, chemical potential, and angular velocity. Varying any of these thermodynamic or kinematic parameters modifies both the in-medium pole positions and the polarization splitting of the spectral functions. The complete parameter dependence therefore cannot be summarized by a few universal monotonic trends. More complete numerical results are presented in Appendix~\ref{app:parameter dependence}, Fig.~\ref{fig:appendix temperature}-\ref{fig:appendix rotation}, and their qualitative behavior is summarized in Tab.~\ref{tab:rho00 trends}.

\begin{table*}[tb]
\caption{Qualitative dependence of the azimuthally averaged spin alignment. The arrows describe the change in $\bar\rho_{00}$ as the indicated parameter increases. ``Low'' and ``high'' $p_t$ refer to the regions on either side of the turning point rather than to universal numerical intervals.}
\label{tab:rho00 trends}
\begin{ruledtabular}
\begin{tabular}{p{0.2\textwidth} p{0.2\textwidth}p{0.2\textwidth}p{0.3\textwidth}}
 Variable& $K^*$ and $\phi$ & $D^*$, $D_s^*$ and $J/\psi$&  Remarks\\
Temperature $T$ & $\uparrow$ at low $p_t$ and $\downarrow$ at high $p_t$& $\downarrow$ at low $p_t$ and $\uparrow$ at high $p_t$& Low-rapidity region ($Y \lesssim 1$)\\
Rapidity $Y$ & $\downarrow$ at low $p_t$ and $\uparrow$ at high $p_t$& $\uparrow$ at low $p_t$ and $\downarrow$ at high $p_t$& Low-rapidity region ($Y \lesssim 1$)\\
Chemical potential $\mu$ & $\uparrow$ at low $p_t$ and $\downarrow$ at high $p_t$& $\downarrow$ at low $p_t$ and $\uparrow$ at high $p_t$& Very small; the absolute change is typically of order $10^{-4}$.\\
Angular velocity $\Omega$ & $\uparrow$ at high $p_t$& No universal trend& The phenomenologically relevant range is $\Omega \leq 0.04~\mathrm{GeV}$.\\
\end{tabular}
\end{ruledtabular}
\end{table*}

Two conclusions are particularly robust. First, varying the parameters reshapes the $p_t$ dependence and can shift the points at which $\bar\rho_{00}-1/3$ or its parameter derivative changes sign. Quoting a trend without specifying the parameter window can therefore be misleading. Second, the response to chemical potential is negligible, while rotation produces a somewhat larger but still modest and nonuniversal modification over the range considered.

\begin{figure*}[!t]
    \centering
    \includegraphics[width=\textwidth]{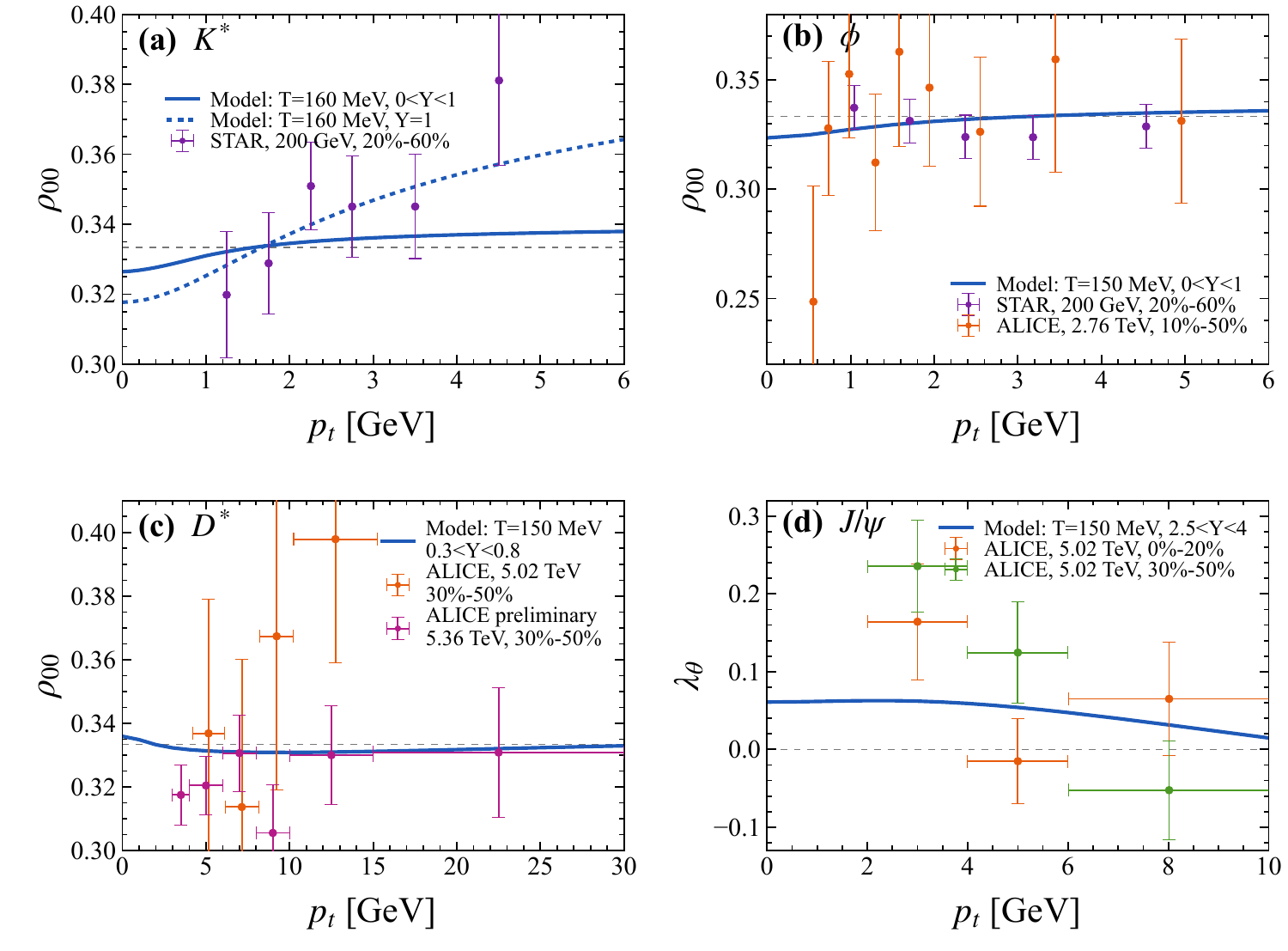}
    \caption{Comparison with representative experimental measurements. (a) $K^*$: the model results at $T=160\,\mathrm{MeV}$ for $0<Y<1$ and $Y=1$ versus STAR Au--Au data at $\sqrt{s_{NN}}=200\,\mathrm{GeV}$~\cite{STAR:2022fan}. (b) $\phi$: the model at $T=150\,\mathrm{MeV}$ and $0<Y<1$ versus STAR and ALICE data~\cite{STAR:2022fan,ALICE:2019aid}. (c) $D^*$: the model at $T=150\,\mathrm{MeV}$ and $0.3<Y<0.8$ compared with the published ALICE Pb--Pb data at $\sqrt{s_{NN}}=5.02\,\mathrm{TeV}$~\cite{ALICE:2025cdf} and the preliminary ALICE measurement in Pb--Pb collisions at $\sqrt{s_{NN}}=5.36\,\mathrm{TeV}$~\cite{ALICE:2026DstarPreliminary}. (d) $J/\psi$: the model prediction for $\lambda_\theta$ at $T=150\,\mathrm{MeV}$ and $2.5<Y<4$ versus ALICE Pb--Pb data at $\sqrt{s_{NN}}=5.02\,\mathrm{TeV}$~\cite{ALICE:2022dyy}. The centrality intervals are specified in the panel legends.}
    \label{fig:experimental comparison}
\end{figure*}

\subsubsection{Comparison with experimental measurements}

Fig.~\ref{fig:experimental comparison} compares the calculated results with experimental data. We set $\Omega=\mu=0$ because the preceding subsection shows that these parameters have only a small effect. For each collision energy, we choose a representative temperature guided by the corresponding chemical-freeze-out systematics~\cite{Flor:2020fdw}. Together with the instantaneous freeze-out prescription, this choice assumes that the spin-density matrix is predominantly fixed near chemical freeze-out and is not substantially modified during the subsequent hadronic evolution.

For $K^*$, the calculated low-$p_t$ suppression and subsequent rise are qualitatively consistent with the STAR result at $\sqrt{s_{NN}}=200\,\mathrm{GeV}$~\cite{STAR:2022fan}. A similar low-$p_t$ suppression of the $\phi$ meson is observed by ALICE~\cite{ALICE:2019aid}. Given the experimental uncertainties, most of the displayed $K^*$ and $\phi$ data points are statistically compatible with the small deviations predicted by the model. However, the much larger $\phi$-meson alignment reported by STAR at lower collision energies is not generated by the model. Lowering the collision energy would mainly increase $\mu$ and slightly change the freeze-out temperature, whereas Tab.~\ref{tab:rho00 trends} shows that the chemical-potential response is too small to account for the observed enhancement.

For $D^*$, the calculation remains close to, but generally below, $1/3$ over most of the transverse-momentum range. It is qualitatively consistent with the published ALICE data at low and intermediate $p_t$, whereas a pronounced discrepancy appears at high $p_t$, where the published results at $\sqrt{s_{NN}}=5.02\,\mathrm{TeV}$ show a substantial enhancement above $1/3$~\cite{ALICE:2025cdf}. Such a discrepancy would not be unexpected if the present holographic framework, which is calibrated primarily to nonperturbative hadronic observables and does not incorporate the perturbative ultraviolet dynamics of QCD, loses descriptive power at large transverse momentum. High-$p_t$ $D^*$ mesons may also receive increasing contributions from hard production, fragmentation, and the long formation times of heavy mesons, all of which lie beyond the scope of the current model. However, the latest preliminary ALICE results at $\sqrt{s_{NN}}=5.36\,\mathrm{TeV}$ differ markedly from the published high-$p_t$ behavior and remain broadly compatible with the calculation throughout the measured range~\cite{ALICE:2026DstarPreliminary}. In particular, the preliminary values below $1/3$ at low and intermediate $p_t$, followed by an approach toward $1/3$ at higher $p_t$, are qualitatively consistent with the model prediction.

For $J/\psi$, Eq.~\eqref{eq:lambda theta} converts the calculated spin alignment into the anisotropy parameter. In the forward-rapidity interval $2.5<Y<4$, the model predicts a positive $\lambda_\theta$ at low $p_t$ that decreases toward zero as $p_t$ increases. This trend is qualitatively consistent with the transverse polarization indicated by the ALICE data~\cite{ALICE:2022dyy}, although the current experimental uncertainties and centrality coverage preclude a stringent quantitative conclusion.

\subsection{Discussion}

The unified spectral-function mechanism accounts for several qualitative observations across flavor sectors, but the calculated deviations generally remain close to the unpolarized value. In particular, the model does not reproduce the strong collision-energy dependence reported for the $\phi$ meson. These discrepancies are not simply failures of the peak-splitting picture; they indicate that equilibrium spectral functions may not fully capture the microscopic mechanisms responsible for spin alignment.

Several microscopic mechanisms have been proposed to explain spin alignment, including thermal vorticity, thermal shear, strong magnetic fields, spin-density fluctuations, and strong-field fluctuations. In principle, these mechanisms can be represented as self-energy corrections to the vector-meson propagator. In the present holographic framework, however, the self-energy is obtained from the response of a classical equilibrium background and therefore represents an effective in-medium self-energy averaged over the thermal ensemble. A mechanism that modifies the average properties of the medium may thus already be encoded, at least implicitly, in the holographic self-energy. By contrast, contributions governed by stochastic fluctuations or higher-order correlation functions are generally absent from a classical holographic background and would require additional fluctuating bulk fields or a more sophisticated construction. In this sense, the present framework provides a well-defined equilibrium baseline.

The thermodynamic parameters are inferred from chemical-freeze-out systematics and serve as a phenomenological bridge between the equilibrium calculation and measurements at different collision energies and centralities. This mapping is useful for identifying robust trends, but it is not a dynamical description. A quantitative comparison will ultimately require coupling the spin-resolved spectral functions to realistic hydrodynamic evolution and channel-dependent production and decay processes.

Within these limitations, the results support a clear physical interpretation: vector-meson spin alignment probes the polarization dependence of the in-medium spectral strength near the vacuum mass shell. The flavor-dependent polarization splitting and the locations of the resulting peaks relative to $m_V$ explain the two qualitative classes. Temperature, rapidity, chemical potential, and rotation reshape this basic on-shell competition without introducing a separate universal mechanism.

\section{Conclusion}
\label{sec:conclusion}

In this work, we have developed a unified phenomenological holographic framework for studying vector mesons across several flavor sectors and applied it to spin alignment in heavy-ion collisions. The model combines an anisotropic Einstein--Maxwell--dilaton background at finite temperature, baryon chemical potential, and angular velocity with a four-flavor soft-wall matter sector. The background reproduces the lattice equation of state and the rotational dependence of the deconfinement temperature. Flavor-dependent infrared mass corrections and auxiliary scalar fields allow the $K^{*}$, $\phi$, $D^{*}$, $D_s^{*}$, and $J/\psi$ mesons to be described within a single framework. The background parameters are constrained by lattice thermodynamics, while the matter-sector parameters are fixed by selected vacuum meson masses.

Within this framework, we have computed the retarded two-point correlation functions of the vector currents and extracted the spectral functions projected onto the three spin-polarization states. The spectral functions contain quasiparticle peaks at the quasinormal frequencies of the bulk vector field. As the temperature rises, these peaks broaden, with the widths of the heavier mesons increasing more slowly. When a meson moves relative to the medium, or when rotation renders the medium anisotropic, the symmetry among the spin states is broken and the longitudinal and transverse spectral functions split. In the present approach, this splitting near the vacuum mass shell is the origin of spin alignment.

The central result is that the five mesons fall into two flavor classes with opposite spin-alignment patterns, as summarized in Fig.~\ref{fig:rho00 overview} and Tab.~\ref{tab:rho00 trends}. For the charmed mesons $D^{*}$, $D_s^{*}$, and $J/\psi$, $\rho_{00}$ exceeds $1/3$ at low transverse momentum and varies nonmonotonically with $p_t$. For the $K^{*}$ and $\phi$ mesons, $\rho_{00}$ is below $1/3$ at low transverse momentum and rises above $1/3$ at intermediate momentum. The two classes also exhibit opposite angular modulations and temperature dependences. Within the present framework, this dichotomy reflects the flavor dependence of the splitting between the longitudinal and transverse in-medium spectral functions near the mass shell. The predictions are qualitatively consistent with several available measurements, including the low-$p_t$ ALICE results for the $K^{*}$ and $\phi$ mesons, the transverse polarization of the $J/\psi$ meson at forward rapidity, and the behavior of the $D^{*}$ meson at low and intermediate $p_t$.

A second robust result is the weak sensitivity of spin alignment to baryon chemical potential and angular velocity over the physically relevant range. Consequently, the enhancement of $\phi$-meson spin alignment with decreasing collision energy observed by STAR cannot be attributed to the equilibrium response of the medium. Likewise, the model does not reproduce the high-$p_t$ $D^{*}$ result, which lies in a kinematic regime where hard production, fragmentation, and perturbative ultraviolet dynamics are expected to become important. These discrepancies delimit the domain of validity of the equilibrium description and provide a baseline for identifying additional microscopic mechanisms.

In the present framework, spin alignment originates from polarization-dependent in-medium self-energy corrections encoded in a classical equilibrium background, which represents an ensemble-averaged response of the strongly coupled medium. This description does not capture contributions governed by stochastic fluctuations or higher-order correlation functions, the nonequilibrium evolution of the fireball before freeze-out, or the perturbative ultraviolet dynamics relevant at high transverse momentum. Natural extensions include fluctuating bulk fields, improved ultraviolet matching, and a dynamical treatment of freeze-out. Together with ongoing and forthcoming measurements for different meson species, this unified description may help clarify the role of the strongly coupled medium in spin alignment.

\begin{acknowledgments}
We acknowledge helpful discussions with Yuanjing Ji, Kun Xu and Guangyu Zheng. This work is supported in part by the National Natural Science Foundation of China (NSFC) Grant Nos: 12235016, 12221005, 12305136, and the start-up funding of Hangzhou Normal University under Grant No. 4245C50223204075. 
\end{acknowledgments}
\appendix
\section{Detailed parameter dependence of the spin alignment}
\label{app:parameter dependence}

This appendix presents the variations of the spin alignment with different values of the relevant parameters, from which the trends summarized in Tab.~\ref{tab:rho00 trends} are derived. Specifically, Figs.~\ref{fig:appendix temperature}–\ref{fig:appendix rotation} show the effects of temperature, rapidity, chemical potential, and angular velocity, respectively, on the $\bar{\rho}_{00}(p_T)$ curves. Here $\bar\rho_{00}$ denotes the azimuthal average defined in Eq.~\eqref{eq:azimuthal average} with $v_2=0.15$.

It is worth noting that pronounced structures localized in transverse momentum appear in the $D^{*}$ and $D_s^{*}$ curves in Fig.~\ref{fig:appendix rotation}. For the parameter set considered here, the longitudinal and transverse peaks of the spin-resolved spectral functions approach and temporarily straddle the vacuum mass shell as $p_t$ increases. Because the two peaks move at different rates, their spectral strengths on the vacuum mass shell vary rapidly. This behavior produces the localized structures in the rotation-induced variation of $\bar\rho_{00}$. The structures therefore arise from the on-shell sampling of rapidly moving spectral peaks rather than from an additional universal rotational effect.

\begin{widetext}
    \centering
    \includegraphics[width=0.96\textwidth]{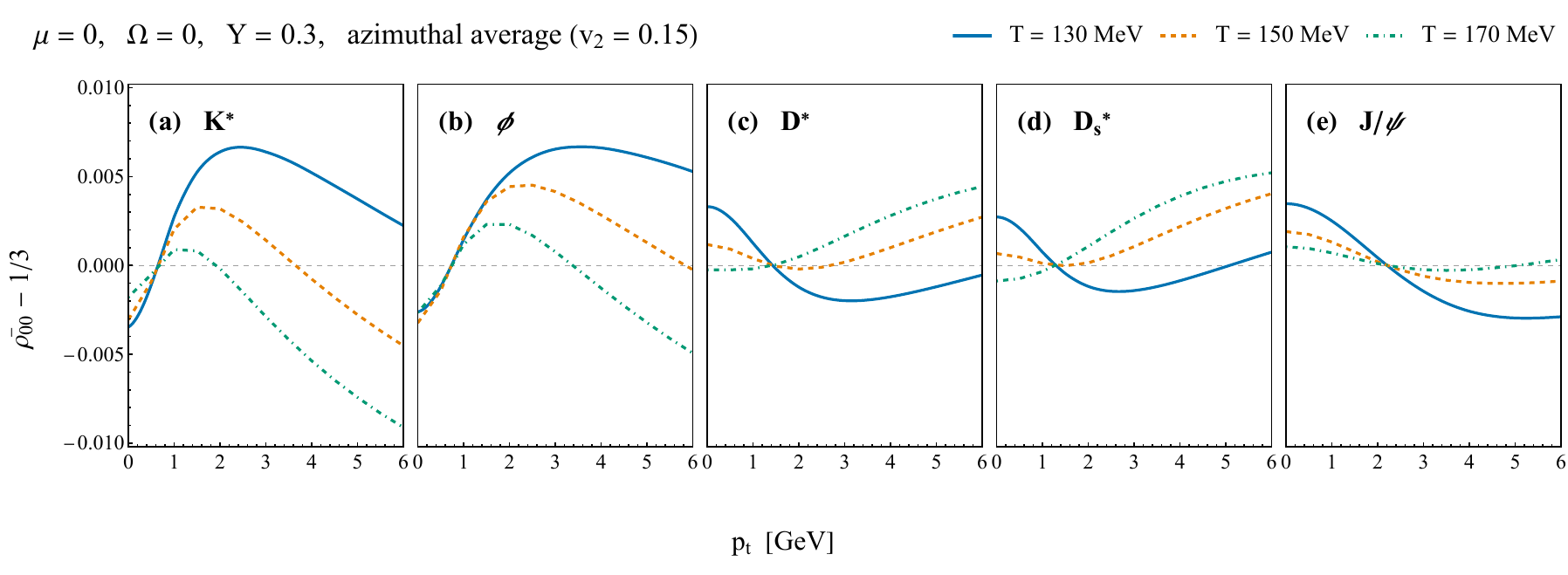}
    \captionof{figure}{Temperature dependence of $\bar\rho_{00}-1/3$ for the five mesons at $\mu=\Omega=0$ and $Y=0.3$. The curves correspond to $T=130$, $150$, and $170\,\mathrm{MeV}$.}
    \label{fig:appendix temperature}
    \vspace{-0.6\baselineskip}
    \includegraphics[width=0.96\textwidth]{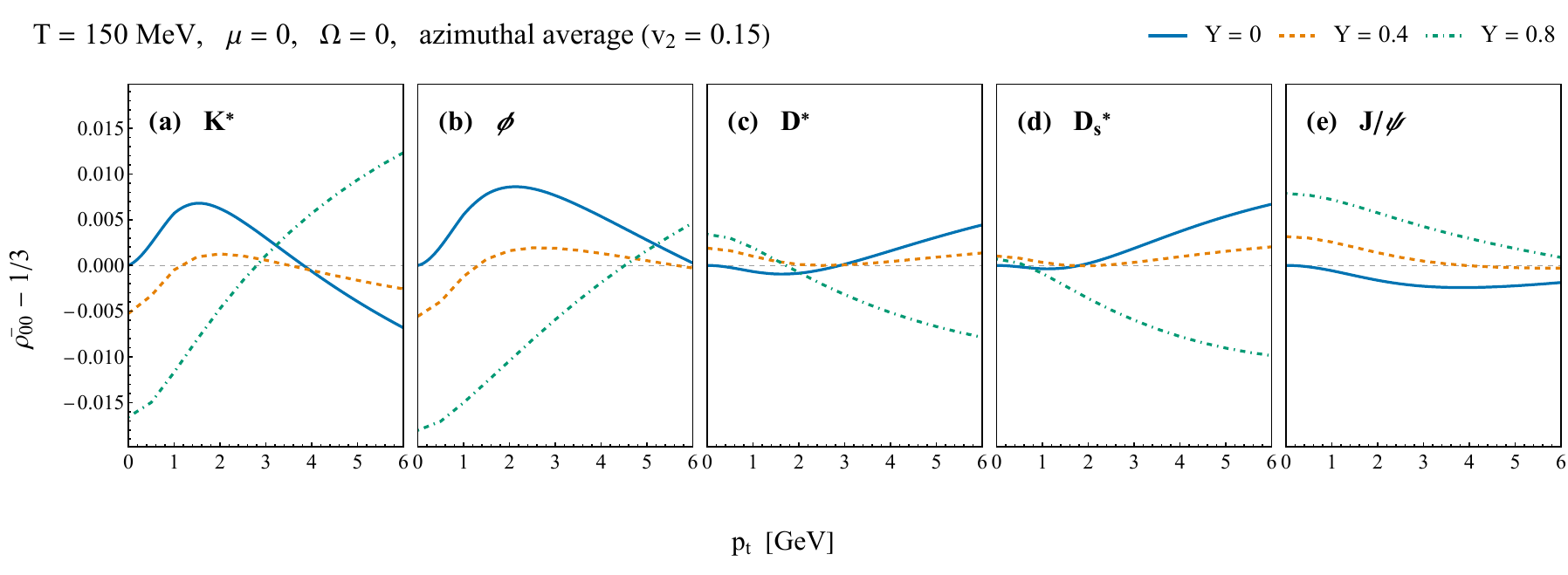}
    \captionof{figure}{Rapidity dependence of $\bar\rho_{00}-1/3$ for the five mesons at $T=150\,\mathrm{MeV}$ and $\mu=\Omega=0$. The curves correspond to $Y=0$, $0.4$, and $0.8$.}
    \label{fig:appendix rapidity}
    
    \includegraphics[width=0.96\textwidth]{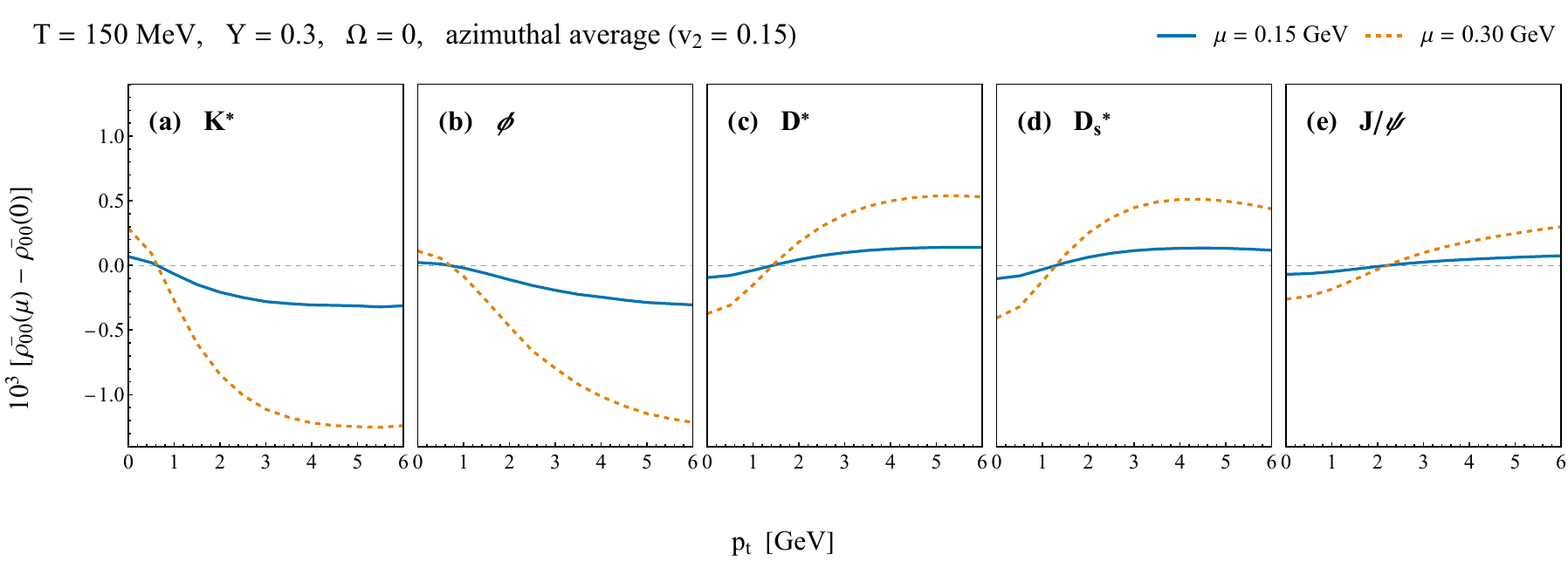}
    \captionof{figure}{Chemical-potential-induced change $10^3[\bar\rho_{00}(\mu)-\bar\rho_{00}(0)]$ for the five mesons at $T=150\,\mathrm{MeV}$, $Y=0.3$, and $\Omega=0$. The curves correspond to $\mu=0.15$ and $0.30\,\mathrm{GeV}$.}
    \label{fig:appendix chemical potential}
    
    \includegraphics[width=0.96\textwidth]{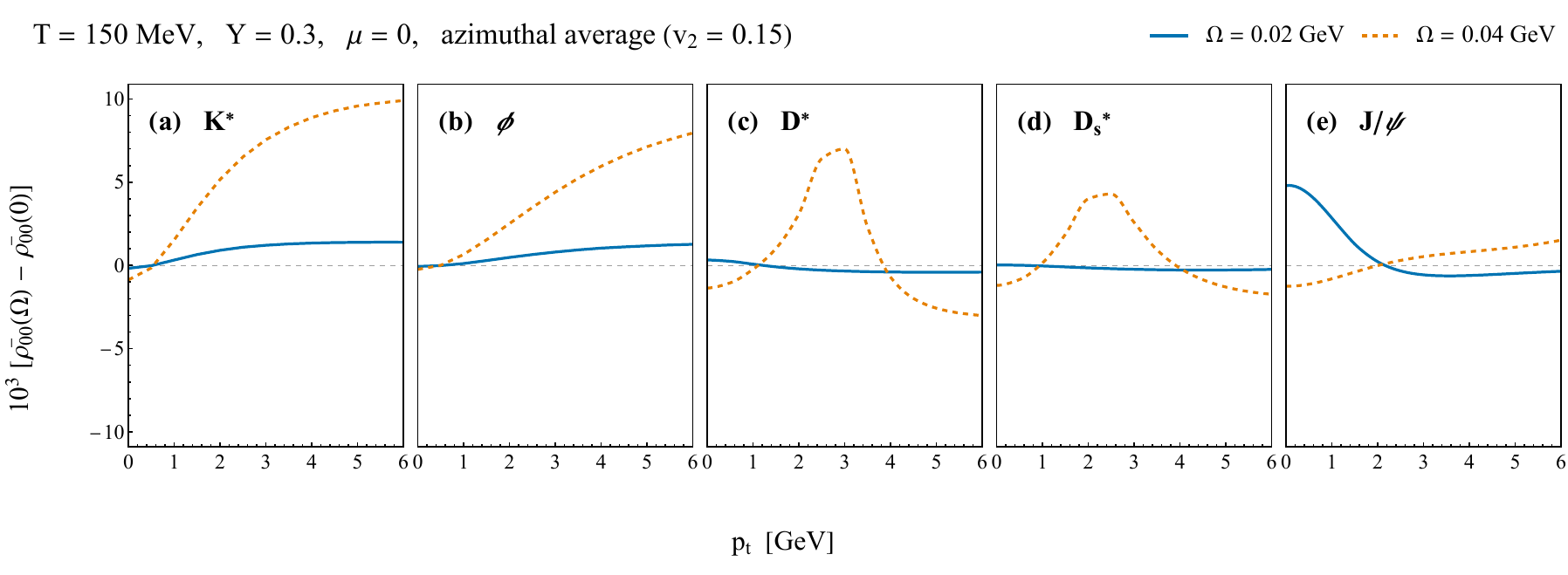}
    \captionof{figure}{Rotation-induced change $10^3[\bar\rho_{00}(\Omega)-\bar\rho_{00}(0)]$ for the five mesons at $T=150\,\mathrm{MeV}$, $Y=0.3$, and $\mu=0$. The curves correspond to $\Omega=0.02$ and $0.04\,\mathrm{GeV}$.}
    \label{fig:appendix rotation}
\end{widetext}

\bibliographystyle{apsrev4-2}
\bibliography{Spin_Alignment_Main}

\end{document}